# Thermodynamic signatures of quantum criticality in cuprates

B. Michon[1,2,3], C. Girod[1,3], S. Badoux[2], J. Kačmarčík[4], Q. Ma[5], M. Dragomir[6],

H. A. Dabkowska[6], B. D. Gaulin[5,6,7], J.-S. Zhou[8], S. Pyon[9], T. Takayama[9], H. Takagi[9],

S. Verret[2], N. Doiron-Leyraud[2], C. Marcenat[10], L. Taillefer[2,7], T. Klein[1,3]

1 Univ. Grenoble Alpes, Institut Néel, F-38042 Grenoble, France

2 Institut quantique, Département de physique & RQMP, Université de Sherbrooke,

Sherbrooke, Québec J1K 2R1, Canada

3 CNRS, Institut Néel, 38042 Grenoble, France

4 Institute of Experimental Physics, Slovak Academy of Sciences, SK-04001 Košice,
Slovakia

5 Department of Physics and Astronomy, McMaster University, Hamilton, Ontario

L8S 4M1, Canada

6 Brockhouse Institute for Materials Research, McMaster University, Hamilton, Ontario

L8S 4M1, Canada

7 Canadian Institute for Advanced Research, Toronto, Ontario M5G 1Z8, Canada

8 University of Texas - Austin, Austin, Texas 78712, USA

9 Department of Advanced Materials, University of Tokyo, Kashiwa 277-8561, Japan

10 Univ. Grenoble Alpes, CEA, INAC, PHELIQS, LATEQS, F-38000 Grenoble, France

**The three central phenomena of cuprate superconductors are linked by a common doping $p^*$, where the enigmatic pseudogap phase ends, around which the superconducting phase forms a dome, and at which the resistivity exhibits an anomalous linear dependence on temperature as $T \to 0$ (ref. 1). However, the** **fundamental nature of $p*$ remains unclear, in particular whether it marks a true quantum phase transition[2]. We have measured the specific heat $C$ of the cuprates Eu-LSCO and Nd-LSCO at low temperature in magnetic fields large enough to suppress superconductivity, over a wide doping range across $p*$ (ref. 3). As a function of doping, we find that the electronic term $C_{el} / T$ is strongly peaked at $p*$, where it exhibits a $\log(1/T)$ dependence as $T \rightarrow 0$. These are the classic signatures of a quantum critical point[4,5,6], as observed in heavy-fermion[7] and iron-based[8] superconductors where their antiferromagnetic phase ends. We conclude that the pseudogap phase of cuprates ends at a quantum critical point, whose associated fluctuations are most likely involved in the $d$-wave pairing and the anomalous scattering.**

In the phase diagram of several organic, heavy-fermion and iron-based superconductors, superconductivity forms a dome around the quantum critical point (QCP) where a phase of antiferromagnetic order ends. The spin fluctuations associated with that QCP are believed to cause both pairing and scattering[5]. The scattering is anomalous in that it produces a resistivity with a linear temperature dependence as $T \rightarrow 0$, instead of the conventional $T^2$ dependence of a Fermi liquid[4,6,9]. In hole-doped cuprates, these same features – a $T_c$ dome and a $T$-linear resistivity – are also observed[1,10], but not at the critical doping where the Néel temperature $T_N$ for the onset of long-range antiferromagnetic order vanishes. Instead, they are observed near the doping $p*$ where the pseudogap phase ends (Fig. 1) and the essential nature of this phase remains unknown – it is the central enigma of cuprates.

The thermodynamic signature of a QCP is a diverging electronic mass. For an antiferromagnetic QCP in two dimensions, for example, the mass is expected to go as $m* \sim \log(1/|x$-$x*|)$ and the specific heat $C$ as $C / T \sim \log(1/|x$-$x*|)$, as one moves the system towards its QCP at $x*$ by varying some tuning parameter $x$, such as pressure or concentration[4]. This is what is observed in the iron-based superconductor $BaFe_2(As_{1-x}P_x)_2$ at its antiferromagnetic QCP (tuned by P concentration), both in the carrier mass measured via quantum oscillations and in the electronic specific heat estimated from the jump at $T_c$ (refs. 6,8). At $x = x*$, one expects that $C / T \sim \log(1/T)$, as observed in the heavy-fermion metal $CeCu_{6-x}Au_x$ at its antiferromagnetic QCP (tuned by Au concentration) (refs. 4,7). This logarithmic divergence of $C / T$ as $T \rightarrow 0$ is the true sign of an energy scale that vanishes at $x^*$.

In the cuprate YBa$_2$Cu$_3$O$_y$ (YBCO), it has long been known that the specific heat jump at $T_c$, $\delta\gamma(T_c)$, decreases dramatically below $p^* \sim 0.19$ (refs. 11,12), in agreement with the decrease in the mass detected via quantum oscillations[13] (Fig. S1). However, it is unclear whether this decrease is due to quantum criticality as $p$ is tuned away from $p^*$ or simply due to a loss of density of states, because a gap is opening. What has been missing is a measurement of the temperature dependence of the normal-state specific heat of a cuprate as $T \to 0$, at $p^*$. This is what we report here.

We have measured $C(T)$ in the closely related cuprate materials La$_{1.8-x}$Eu$_{0.2}$Sr$_x$CuO$_4$ (Eu-LSCO) and La$_{1.6-x}$Nd$_{0.4}$Sr$_x$CuO$_4$ (Nd-LSCO). Because of their low $T_c$ ($< 20$ K), superconductivity can be fully suppressed with a readily accessible magnetic field ($\sim 15$ T). The two materials have the same crystal structure and phase diagram, with very similar boundaries for the pseudogap phase, $T^*(p)$ (Fig. 1), and superconducting phase, $T_c(p)$ (Fig. S2). In Nd-LSCO, resistivity and Hall effect measurements yield $p^* = 0.23 \pm 0.01$ (ref. 3). At $p = 0.24$, $\rho(T)$ is linear in $T$ as $T \to 0$ (refs. 3,10), the signature of quantum criticality in electrical transport. At $p < 0.23$, $\rho(T)$ exhibits a pronounced upturn at low $T$ (refs. 3,10), due to the loss of carrier density caused by the opening of the pseudogap below $T^*$ (refs. 3,14). A very similar behavior is observed in Eu-LSCO at $p = 0.24$ and $p = 0.21$, respectively (Fig. S3). Nernst measurements reveal that Eu-LSCO and Nd-LSCO have the same $T^*$ at $p = 0.21$ (ref. 15) (Fig. 1). We deduce that $p^* \sim 0.23$ also in Eu-LSCO.

The specific heat of 5 crystals of Eu-LSCO and 7 crystals of Nd-LSCO was measured below 10 K (Fig. S4). The normal-state specific heat is obtained by applying a magnetic field of either $H = 8$ T or 18 T (Figs. S5 and S6). In Figs. 2a and 3b, we show normal-state data at $H = 18$ T in Eu-LSCO and Nd-LSCO, respectively, plotted as $C / T$ vs $T^2$. Note that the magnetic moment on the Nd produces a Schottky anomaly in the specific heat of Nd-LSCO, not present in Eu-LSCO (since Eu has no moment). At $H = 0$, this anomaly leads to a large increase of $C / T$ at low $T$ (Fig. 3a). However, a magnetic field moves this Schottky contribution to higher temperature, so that it becomes negligible below $\sim 5$ K at 18 T (Fig. 3a).

In Fig. 4a, we plot the raw data at $T = 2$ K (and $H = 18$ T), $C / T$ vs $p$, for all 12 crystals.

The 12 data points fall on the same smooth curve, demonstrating a high level of quantitative fidelity and reproducibility. Because the magnetic and nuclear (see below) Schottky contributions are both negligible at 2 K, we have $C = C_{el} + C_{ph}$, the sum of electronic and phononic contributions. In Eu-LSCO at $p = 0.11$ and $p = 0.16$, the data obey $C / T = \gamma + \beta T^2$ below $T \sim 5$ K (Fig. 2a). The residual linear term $\gamma$ is electronic and the second term is due to phonons, with $\beta \sim 0.22$ mJ / K$^4$ mol for both dopings. The same is true in Nd-LSCO at $p = 0.12$ (Fig. 3a) and $p = 0.15$ (Fig. S4c), again with $\beta \sim 0.22$ mJ / K$^4$ mol for both dopings. In Fig. S7d, we plot $\beta$ vs $p$ for our various samples.

We have measured the doping dependence of the low-energy phonon density of states with neutrons, and found that the peak energy, $E_{ph}$, increases slightly with doping at first and then saturates at high doping (Fig. S8). The observed $p$ dependence of $E_{ph}$ is consistent with our full set of data for $\beta$ vs $p$, given that $\beta \sim E_{ph}^{-3}$ (Fig. S7d). Quantitatively, $\beta \sim E_{ph}^{-3}$ decreases by only 7% in going from $p = 0.16$ to $p = 0.24$. Given that $C_{ph} / T = \beta T^2$ at 2 K is 20 times smaller than the measured $C / T$ at $p = 0.24$, and essentially constant vs $p$ (Fig. 4a), the huge rise in $C / T$ up to $p^*$ is unambiguously and entirely due to the electronic term $C_{el} / T$. Of course, this drop in $C_{el} / T$ below $p^*$ is a well-known signature of the pseudogap phase[12].

To investigate what happens above $p^*$, we measured the specific heat of three Nd-LSCO samples, with $p = 0.27$, 0.36 and 0.40 – dopings at which the material is no longer superconducting. Because it is difficult to grow single crystals at such high doping, the samples are polycrystalline powder. As a result, a field cannot be used to shift the magnetic Schottky anomaly ($C_{mag}$) up in temperature (because the effect of a field is highly anisotropic). Nevertheless, we can reliably subtract $C_{mag}$ from $C$ and obtain $C_{el} + C_{ph}$, as demonstrated for $p = 0.12$ in Fig. S9a. In Fig. S9b, we see that powder data at $p = 0.12$ and $H = 0$ obey $(C - C_{mag})/T = \gamma + \beta T^2$ with the same values of $\gamma$ and $\beta$ as those obtained from single-crystal data for $C / T$ at $p = 0.12$ and $H = 18$ T (Fig. 4 and Fig. S7d). Plotting the values of $(C - C_{mag})/T$ at $T = 2$ K in Fig. 4a, we see that $C / T$ drops by a factor 3 in going from $p^*$ up to $p = 0.4$. This drop is nicely consistent with published data on La$_{2-x}$Sr$_x$CuO$_4$ (LSCO) at $p = 0.33$ (ref. 16) (Fig. 4a).

The thermodynamic signature of the pseudogap critical point in Eu-LSCO and Nd-LSCO

is therefore seen to be a huge peak in $C_{el}$ / $T$ at $p^*$, not just a drop below $p^*$. There are two standard explanations for such a peak: a van Hove singularity (vHs) in the band structure and a quantum critical point. Because there is indeed a vHs in Nd-LSCO at $p = p_{vHs} \sim p^*$ (ref. 17), we have considered the first scenario carefully. Fortunately, the band structure of Nd-LSCO is well known and very simple[17], so reliable calculations can be performed[18]. In a perfectly clean two-dimensional metal, $C_{el}$ / $T$ vs $p$ does show a sharp cusp at $p = p_{vHs}$ in the $T = 0$ limit (Fig. S10). However, when the actual 3D dispersion of the Fermi surface and the actual level of disorder scattering are included, the peak due to the vHs is dramatically reduced and broadened (Fig. S10). The large and sharp peak we observe is therefore clearly due to electronic effects beyond the band structure.

We now turn to the temperature dependence of $C_{el}$. In Fig. 2a, we see that $C$ / $T$ at $p = 0.24$ deviates strongly from the $\gamma + \beta T^2$ behavior as $T \to 0$. To investigate this deviation in detail, measurements were carried out in a $^3$He refrigerator with a field of 8 T, just enough to reach the normal state at $p = 0.16$ and 0.24 (Fig. S6). The data are plotted as $C$ / $T$ vs $T$ in Fig. 2b. Below 1 K, we observe a nuclear Schottky anomaly ($C_{nuclear}$), which rises as $C$ / $T \sim T^{-3}$. In Fig. 2c, we plot the difference $\Delta C$ / $T = C$ / $T - \gamma$, *i.e.* the raw data minus a constant, on a log-log plot. For $p = 0.11$ and $p = 0.16$, $\gamma = 2.8$ and 4.2 mJ /K$^2$ mol, respectively, obtained from a fit to $C$ / $T = \gamma + \beta T^2$ below 5 K (see above). We see that $\Delta C$ / $T$ is the same at $p = 0.11$ and $p = 0.16$, over the entire range from 0.5 K to 10 K. This shows that the Schottky contribution and the phonon background do not change detectably when increasing $p$ from 0.11 to 0.16. It also shows that the electronic specific heat $C_{el}(T) = \gamma\, T$ up to 10 K, at those two dopings. In order to obtain the $T$ dependence of $C_{el}(T)$ at other dopings, we can then simply subtract $\Delta C$ measured in the $p = 0.16$ sample from the raw data of each sample, *i.e.* $C_{el} = C(p; T) - \Delta C(p{=}0.16; T)$. (In other words, from each curve in Fig. 2b, we subtract the curve at $p = 0.16$, and add 4.2 mJ /K$^2$ mol.) The result is plotted as $C_{el}$ /$T$ vs log$T$ in Fig. 2d, where we see that $C_{el}$ /$T$ at $p = 0.24$ is linear from $\sim$ 10 K down to 0.5 K, our lowest temperature. Note that allowing for a 7% decrease in $C_{ph}$ from $p = 0.16$ to $p = 0.24$, consistent with the neutron data (Fig. S8), does not affect the log$T$ dependence below 5-6 K (Fig. S7c). In summary, we arrive at the robust finding that $C_{el}$ /$T \sim$ log$(1/T)$ at $p \sim p^*$.

In Fig. 3b (and Fig. S4c), we show Nd-LSCO data obtained in the $^3$He refrigerator at $H = 8$ T, a field sufficient to suppress superconductivity at $p = 0.12$, $0.22$, $0.23$, $0.24$ and $0.25$. As was done for Eu-LSCO, we obtain $C_{el}(T)$ by subtracting a reference curve at low doping, in this case at $p = 0.12$, from the raw data at all other dopings. The result is displayed in Fig. 3d (and Fig. S4d), plotted as $C_{el}/T$ vs $\log T$. The Nd-LSCO data for $C_{el}(T)$ (Fig. 3d) are seen to be in excellent quantitative agreement with the Eu-LSCO data (Fig. 2d). In particular, they confirm our key finding that $C_{el}/T \sim \log(1/T)$ at $p \sim p^*$. We therefore find that the cuprates Eu-LSCO and Nd-LSCO exhibit the classic thermodynamic signature of a QCP, as observed in the antiferromagnetic heavy-fermion metals CeCu$_{6-x}$Au$_x$ (ref. 7), YbRh$_2$Si$_2$ (ref. 19), and CeCoIn$_5$ (ref. 20). By contrast, the contribution of the vHs to $C_{el}/T$ is completely flat in temperature below 10 K, because of the significant 3D dispersion and disorder (Fig. S10). Note that strong disorder does not alter the $\log(1/T)$ dependence of $C_{el}/T$ coming from a QCP, as demonstrated by its persistence down to $\sim 20$ mK in samples of CeCu$_{5.9}$Au$_{0.1}$ (ref. 7) that have a residual resistivity $\rho_0$ larger than that of our own samples at $p = 0.24$ (Fig. S3).

In Fig. 4b, we plot the value of $C_{el}/T$ for our 12 crystals as a function of doping, estimated at three temperatures: 0.5 K, 2 K, and 10 K. We also plot the extrapolated $\gamma$ values for our five polycrystalline samples of Nd-LSCO ($x = 0.07$, $0.12$, $0.27$, $0.36$ and $0.40$; Fig. S9b) and the LSCO crystal at $x = 0.33$ (ref. 16). The resulting curve at 2 K is essentially identical to the raw data at 2 K (Fig. 4a), confirming the validity of our subtraction procedure to extract $C_{el}$ from the measured $C$. Taken together, the $p$ and $T$ dependences of $C_{el}$ provide compelling evidence for a QCP in Eu-LSCO and Nd-LSCO.

The strong similarity of our data on Nd-LSCO and Eu-LSCO with data on other cuprates indicates that the signatures of a QCP reported here are likely to be a generic feature of hole-doped cuprates. In Fig. S11, we compare, across the full doping range, our values of $C_{el}/T$ in Eu-LSCO and Nd-LSCO at 10 K with $\gamma$ values in LSCO obtained from fits to $C/T = \gamma + \beta T^2$ between $\sim 4$ K and $\sim 8$ K, where $C$ was measured on LSCO powders made non-superconducting by adding high levels of Zn impurities[21]. (Note that in this early work the temperature dependence of $C_{el}/T$ was not investigated at very low temperature.) A clear peak in $\gamma$ vs $p$ is observed also in LSCO, similar to that found here in Nd-LSCO, although centered at a lower doping – consistent with the fact that $p^*$ is lower in that material (Fig. 1).

In no other cuprate has a direct measurement of the normal-state specific heat at low temperature been performed across $p^*$. We must therefore piece together different data from different materials. This is what we do in Fig. S1b. Starting on the overdoped side, we have $\gamma = 6.6 \pm 1$ mJ / $K^2$ mol in $Tl_2Ba_2CuO_{6+\delta}$ (Tl-2201) at $p \sim 0.35$ (ref. 22), not far from $\gamma = 7.6 \pm 0.6$ mJ / $K^2$ mol obtained from the effective mass measured by quantum oscillations, $m^* = 5.2 \pm 0.4\ m_e$, in Tl-2201 at $p = 0.3$ (ref. 23). Since those are similar to the values found in Nd-LSCO and in LSCO at $p \sim 0.3$ (Fig. S1a), it is reasonable to suppose that other cuprates, such as YBCO, $Bi_2Sr_2CaCu_2O_8$ and $HgBa_2CuO_{4+\delta}$ (Hg-1201), would also have $\gamma \sim 7$ mJ / $K^2$ mol (and $m^*$) at $p \sim 0.3$. (None of these 3 materials has been measured beyond $p \sim 0.23$.) On the underdoped side, quantum oscillations in YBCO (ref. 13) and Hg-1201 (ref. 24) at $p \sim 0.1$ yield $\gamma = 2.5$ and 4.0 mJ / $K^2$ mol, respectively (per mole of planar Cu), compared to $\gamma = 2.8$ and 3.6 mJ / $K^2$ mol in Eu-LSCO at $p = 0.11$ and Nd-LSCO at $p = 0.12$, respectively (Fig. 4b) – all in good agreement. We emphasize that $\gamma$ in Tl-2201 at $p \sim 0.35$ is only a factor 1.7 larger than $\gamma$ in Hg-1201 at $p \sim 0.1$. In other words, the opening of the pseudogap between the two has only reduced the density of states by a factor $\sim 1.7$. This is a much smaller reduction than that observed in going from $p^*$ to $p = 0.1$ in any hole-doped cuprate[11,12,25], and therefore $\gamma$ in Tl-2201 must first rise from $p \sim 0.35$ to $p^*$ before it falls below $p^*$. In other words, we argue that $\gamma$ must go through a peak at $p^*$ in cuprates quite generally.

In Fig. S1b, we plot the specific heat jump at $T_c$, $\delta\gamma(T_c)$, as a function of $p$, previously measured in YBCO (ref. 12). We see that $\delta\gamma(T_c)$ drops by a factor $\sim 10$ in going from $p^*$ to $p \sim 0.1$. Since $\delta\gamma(T_c) \sim \gamma$, this implies that $\gamma \sim 25$ mJ / $K^2$ mol at $p^*$ in YBCO, comparable to our value of $C_{el} / T = 22$ mJ / $K^2$ mol at $T = 0.5$ K in Nd-LSCO at $p^*$. This high value of $\gamma$ in YBCO must then drop by a factor $\sim 3$-4 above $p^*$, to reach the common value $\gamma \sim 7$ mJ / $K^2$ mol at $p \sim 0.3$ (Fig. S1b). Our finding that $\gamma$ (and $m^*$) peaks at $p^*$ is a change of paradigm in our understanding of cuprates – it reveals a mechanism of strong mass enhancement above $p^*$, associated with a QCP at $p^*$.

Our observation of a continuous logarithmic increase of the electronic specific heat down to temperatures as low as 0.5 K raises the fundamental question of what energy scale, associated with most of the low-temperature entropy, vanishes at $p^*$ ? And what corresponding length scale diverges at $p^*$ ?

Given the remarkable similarity with the signatures of an antiferromagnetic QCP (ref. 26), as observed in heavy-fermion metals[7] and iron-based superconductors[8], it is tempting to attribute the quantum criticality of cuprates to AF spin correlations. However, unlike in electron-doped cuprates where the AF correlation length diverges as $T \to 0$ (ref. 27) close to the critical doping $x^* \sim 0.16$ where the Fermi surface is reconstructed[28] and $\rho(T)$ is linear down to 40 mK (ref. 29), there is no evidence of a diverging AF correlation length in hole-doped cuprates[30].

In Nd-LSCO, incommensurate spin-density-wave (SDW) order is observed by neutron diffraction[31] to vanish with increasing $p$ at $p \sim p^*$, but there is no divergent SDW correlation length or vanishing energy scale. Indeed, muon spin relaxation finds that the magnetic order in Nd-LSCO vanishes before $p^*$, at $p \sim 0.20$, which indicates that the order is not static at $p = 0.20$ (ref. 32), or above.

It may therefore be that the quantum criticality of hole-doped cuprates is of an entirely new kind, possibly involving topological order[33]. Alternatively, we may be dealing with two closely intertwined mechanisms: one mechanism for the pseudogap, possibly short-range AF correlations, and a second – separate but coupled – mechanism for the quantum criticality, possibly nematic order[34] or current-loop order[35].

**Acknowledgements.** C.M. and T.K. acknowledge support from the Laboratoire d'excellence LANEF (ANR-10-LABX-51-01) and the Laboratoire National des Champs Magnétiques Intenses (LNCMI) in Grenoble. L.T. acknowledges support from the Canadian Institute for Advanced Research (CIFAR) and funding from the Natural Sciences and Engineering Research Council of Canada (NSERC; PIN:123817), the Fonds de recherche du Québec - Nature et Technologies (FRQNT), the Canada Foundation for Innovation (CFI), and a Canada Research Chair. This research was undertaken thanks in part to funding from the Canada First Research Excellence Fund. Part of this work was funded by the Gordon and Betty Moore Foundation's EPiQS Initiative (Grant GBMF5306 to L.T.). J.-S.Z. was supported by DOD-ARMY (W911NF-16-1-0559) in the US. H.T. acknowledges MEXT Japan for a Grant-in-Aid for Scientific Research.

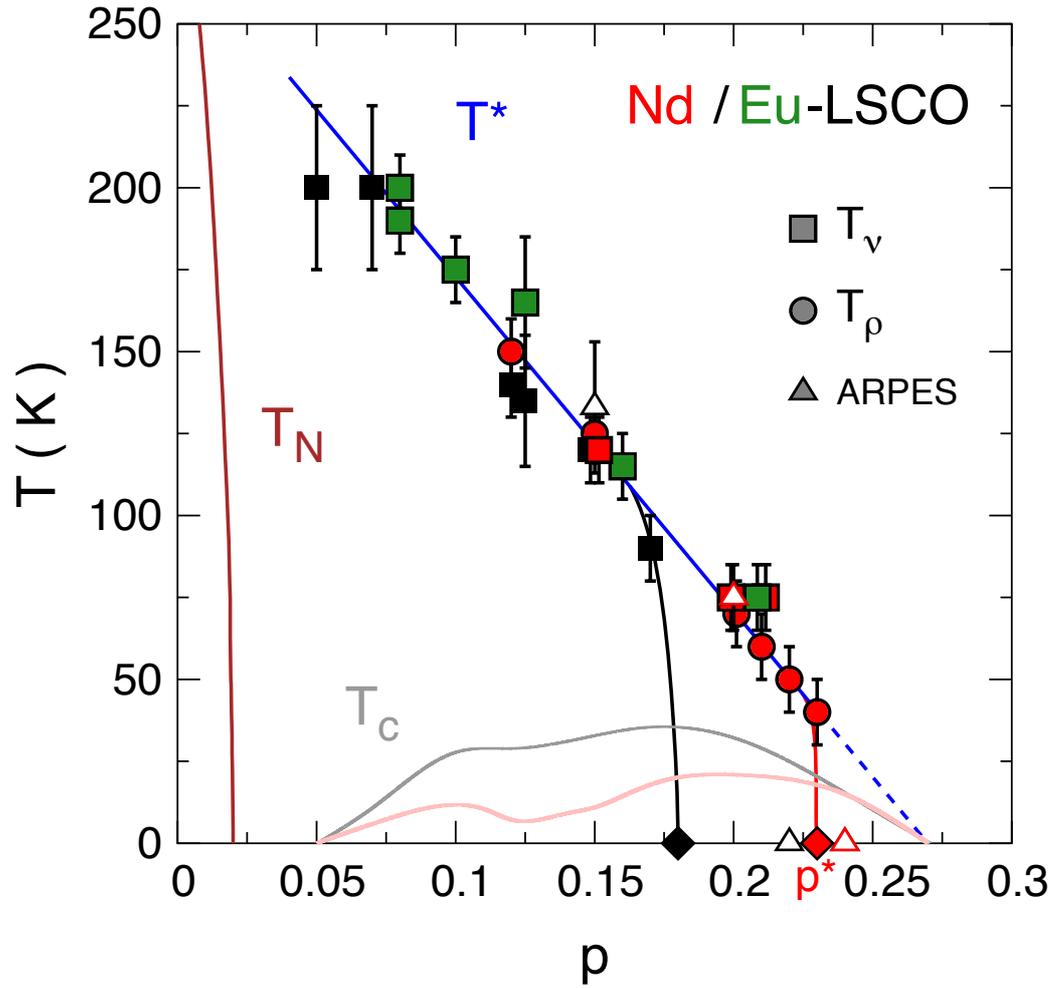

**Fig. 1 | Temperature-doping phase diagram of LSCO-based cuprates.**

Temperature-doping phase diagram of LSCO (black), Nd-LSCO (red) and Eu-LSCO (green), showing the boundary of the phase of long-range commensurate antiferromagnetic order ($T_N$, brown line), the pseudogap temperature $T^*$ (blue line) and the superconducting transition temperature $T_c$ of LSCO (grey line) and Nd-LSCO (pink line). $T^*$ is detected in two transport properties : resistivity ($T_\rho$, circles) and the Nernst effect ($T_\nu$, squares). The open triangles show $T^*$ detected by ARPES as the temperature below which the anti-nodal pseudogap opens, in LSCO (black) and Nd-LSCO (red). We see that $T_\nu \simeq T_\rho \simeq T^*$, within error bars. The pseudogap phase ends at a critical doping $p^* = 0.18 \pm 0.01$ in LSCO (black diamond) and $p^* = 0.23 \pm 0.01$ in Nd-LSCO (red diamond). Figure adapted from ref. 15.

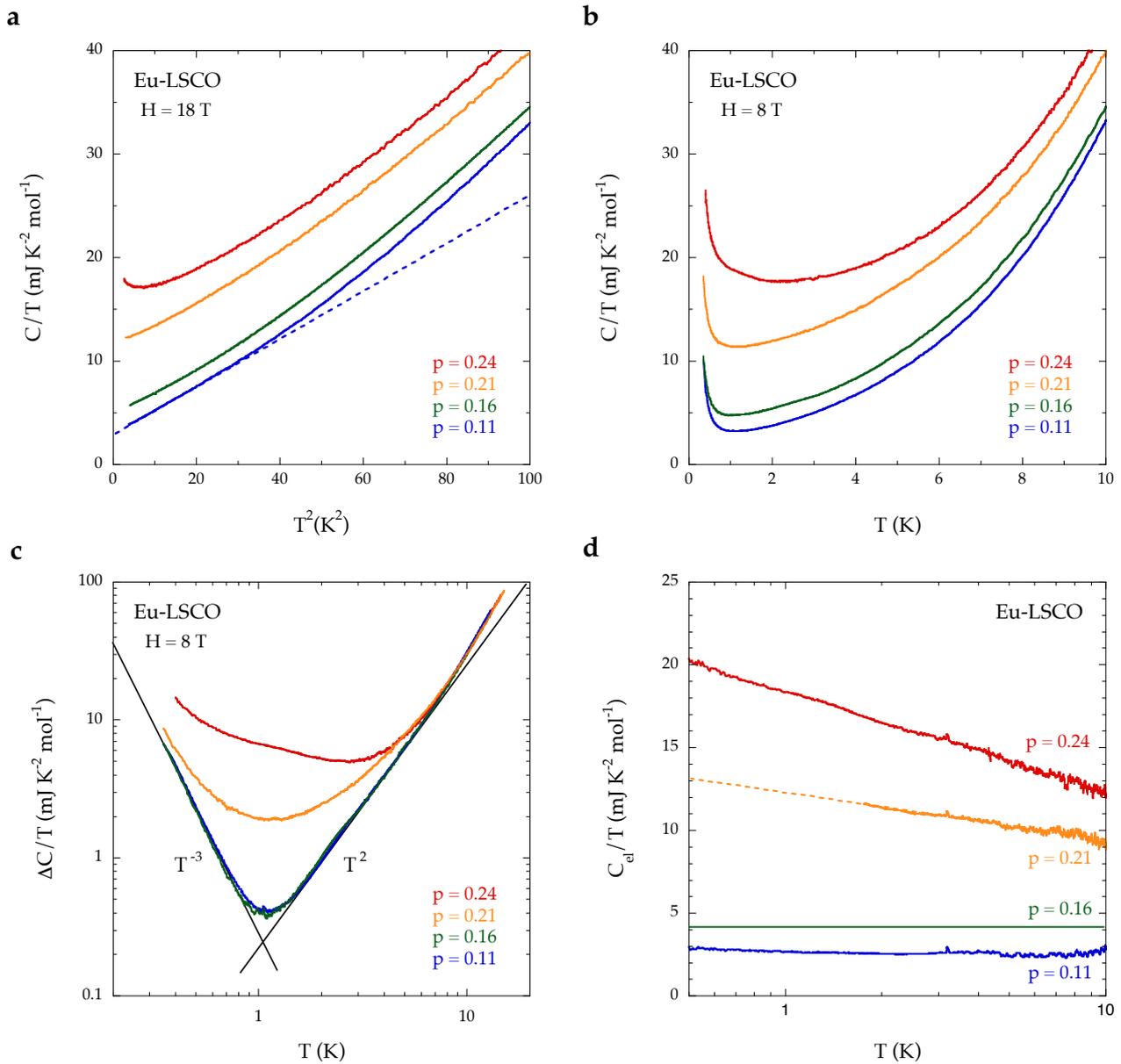

**Fig. 2 | Specific heat of Eu-LSCO.**

**a)** Specific heat of Eu-LSCO measured in a field $H$ = 18 T, plotted as $C$ / $T$ vs $T^2$, for four different dopings, as indicated. The dashed line is a linear fit to the data at $p$ = 0.11 (blue) for $T <$ 5 K; it yields $\gamma$ = 2.8 mJ /K$^2$ mol and $\beta$ = 0.22 mJ / K$^4$ mol, where $C$ / $T$ = $\gamma$ + $\beta T^2$. (The same fit to data at $p$ = 0.16 (green) yields $\gamma$ = 4.2 mJ /K$^2$ mol and $\beta$ = 0.22 mJ / K$^4$ mol.) **b)** Specific heat of the same samples measured in a field $H$ = 8 T, down to 0.4 K. The rapid rise below 1 K is a nuclear Schottky anomaly. **c)** Difference between the measured $C$ / $T$ of panel b and a constant term $\gamma$, plotted for each doping as a function of temperature, on a log-log plot ($\gamma$ = 2.8 and 4.2 mJ /K$^2$ mol, at $p$ = 0.11 and 0.16, respectively). The line marked $T^2$ shows that the data at $p$ = 0.11 and $p$ = 0.16 obey $\Delta C = \beta T^3$ in the range from 1.5 K to ~ 5 K. The line marked $T^{-3}$ shows that the data at $p$ = 0.11 and $p$ = 0.16 obey $\Delta C \sim T^{-2}$ below 1 K, as expected for the upper tail of a Schottky anomaly. The $\Delta C$ curve at $p$ = 0.16, $\Delta C(p$=0.16; $T)$, therefore constitutes the non-electronic, and weakly doping-dependent, background for $C(T)$ in Eu-LSCO, made of phonon and Schottky contributions. **d)** Electronic specific heat $C_{\text{el}}(T)$, defined as $C(p; T) - \Delta C(p$=0.16; $T)$, plotted as $C_{\text{el}} /T$ vs log$T$, from data at $H$ = 8 T ($p$ = 0.11, 0.16, and 0.24) and at $H$ = 18 T ($p$ = 0.21). (The dashed line is a linear extrapolation of the $p$ = 0.21 data.) At $p$ = 0.11, $C_{\text{el}} /T = \gamma$, a constant, while at $p$ = 0.24 ~ $p$*, $C_{\text{el}} /T \sim \log(1/T)$, the thermodynamic signature of a quantum critical point. (See Fig. S4 for the complete set of dopings, and Fig. S7 for further analysis and discussion.)

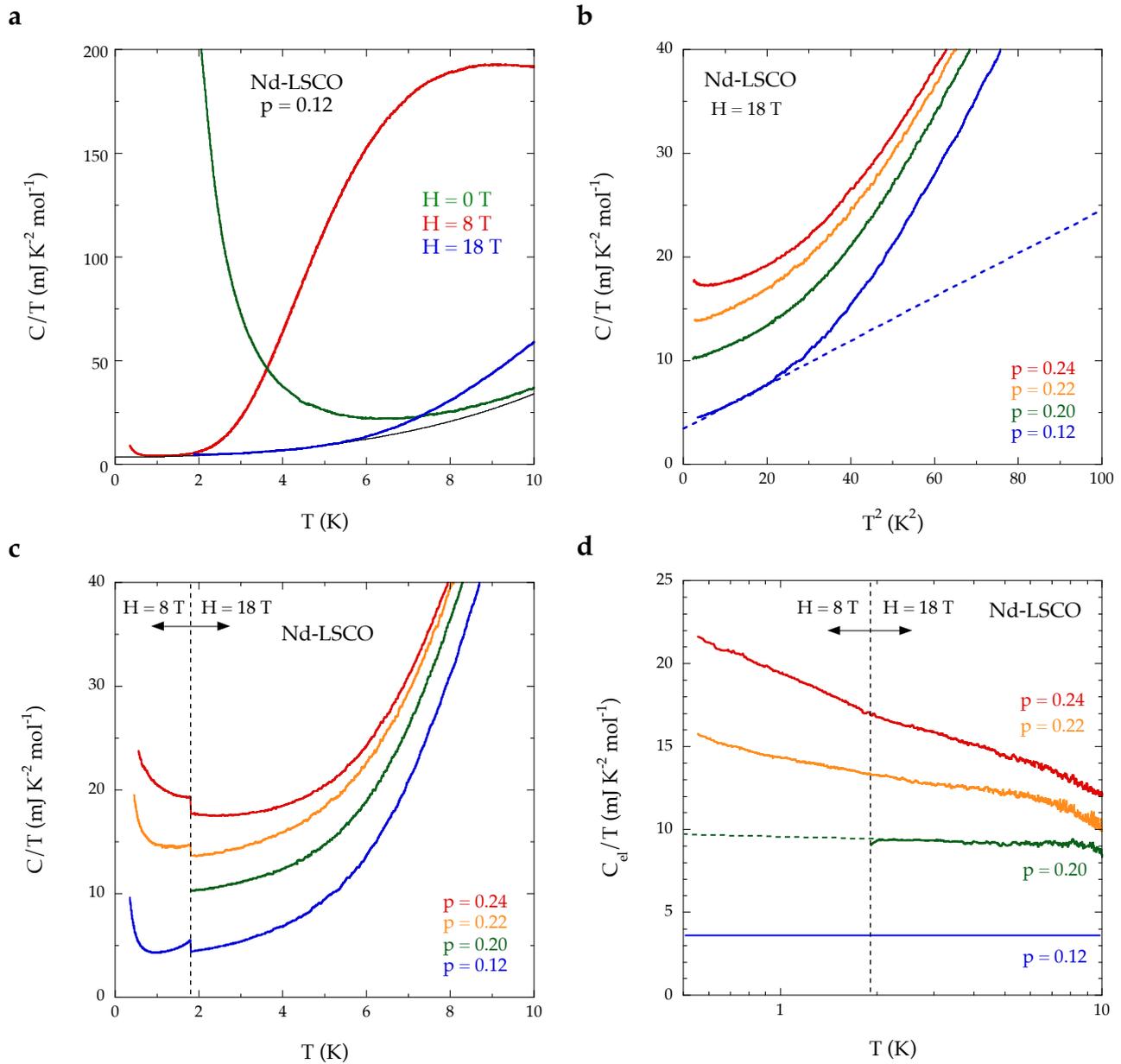

**Fig. 3 | Specific heat of Nd-LSCO.**

**a)** Specific heat of our Nd-LSCO crystal with $p = 0.12$, plotted as $C / T$ vs $T$ at three different fields, as indicated. At $H = 0$ (green), we see the large Schottky anomaly associated with Nd ions, varying as $C_{mag} \sim T^{-2}$ at low $T$. At $H = 8$ T (red), it is pushed up above 2 K; at $H = 18$ T (blue), above 5 K. The line is a fit of the 18 T data to $\gamma + \beta T^2$ below 5 K. **b)** Specific heat of Nd-LSCO measured in a field $H = 18$ T, plotted as $C / T$ vs $T^2$, for four dopings, as indicated. (Data for our 7 crystals of Nd-LSCO are displayed in Fig. S4.) The dashed line is a linear fit to the data at $p = 0.12$, $C / T = \gamma + \beta T^2$ (below 5 K), giving $\gamma = 3.6$ mJ /K$^2$ mol and $\beta = 0.215$ mJ / K$^4$ mol. **c)** Same data as in panel b, plotted as $C$ / $T$ vs $T$. Below the vertical dashed line, we show low-temperature data taken at $H = 8$ T on three of these same samples. **d)** Electronic specific heat $C_{el}(T)$, defined as $C(p;\ T) - \Delta C(p=0.12;\ T)$, plotted as $C_{el}$ /$T$ vs $\log T$ (see Fig. 2). This is done separately for the data below and above the dashed line in panel c. At $p = 0.24 \sim p^*$, $C_{el}$ /$T \sim \log(1/T)$, just as in Eu-LSCO (Fig. 2d).

**a**

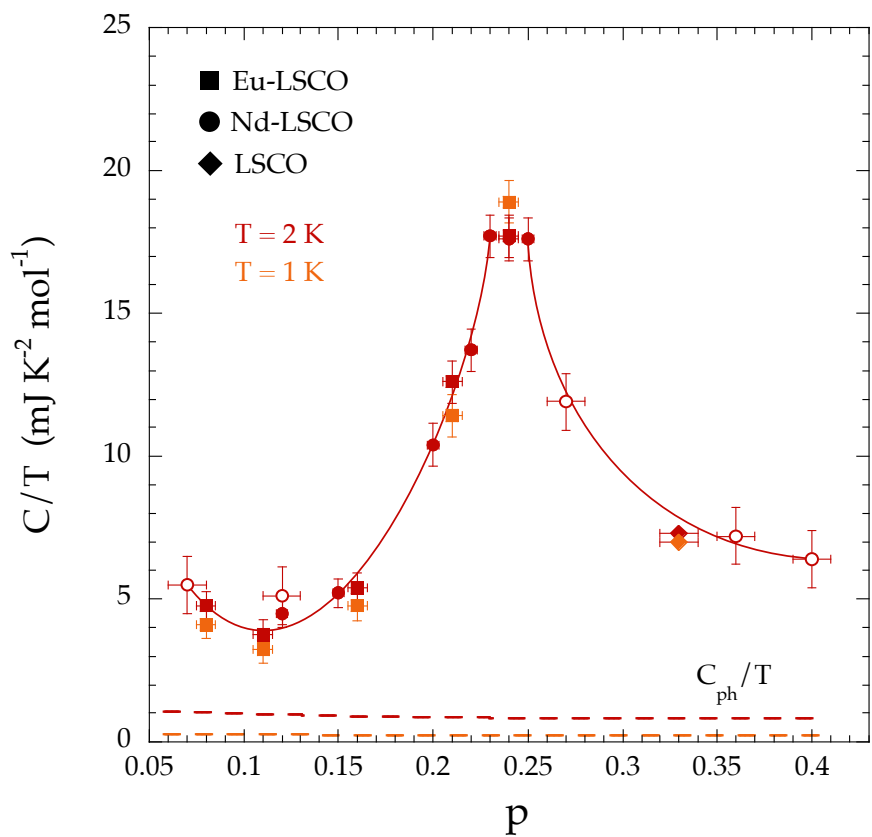

**b**

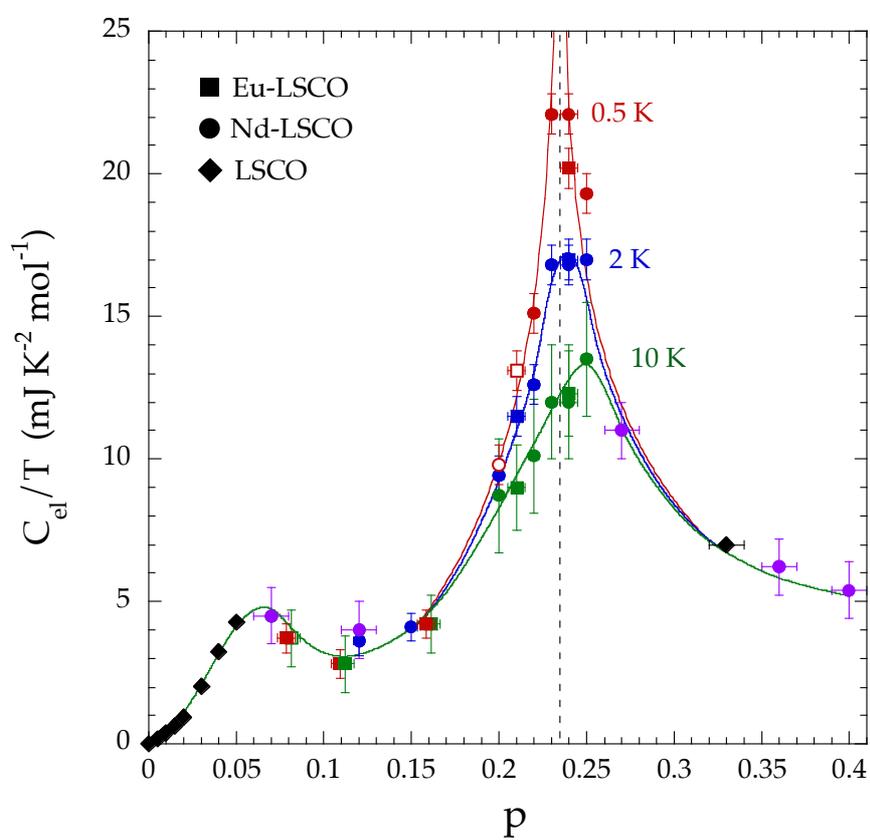

**Fig. 4 | Doping dependence of the electronic specific heat.**

**a)** Raw data for $C/T$ in Eu-LSCO (squares) and Nd-LSCO (circles) at $T = 2$ K and $H = 18$ T (full red symbols) and at $T = 1$ K and $H = 8$ T (full orange squares). We also include data points for non-superconducting LSCO at $H = 0$ (diamonds), at $T = 2$ K (red) and $T = 1$ K (orange), at $p = 0.33$ (ref. 16). Open circles are $(C - C_{mag})/T$ at $T = 2$ K obtained on polycrystalline samples of Nd-LSCO, as described in Fig. S9. The dashed lines indicate the phonon contribution, $C_{ph}/T = \beta T^2$, to the specific heat of Nd-LSCO, at $T = 2$ K (red) and $T = 1$ K (orange) (Fig. S7). (The phonon term is similar in Eu-LSCO (Fig. S7), and slightly smaller in LSCO.) **b)** Normal-state electronic specific heat $C_{el}$ of Eu-LSCO (squares; from Fig. 2d) and Nd-LSCO (circles; from Fig. 3d), at $T = 0.5$ K (red), 2 K (blue) and 10 K (green), plotted as $C_{el}/T$ vs $p$. (At $p = 0.08$, 0.11 and 0.16, the red and green squares are split apart slightly so they can both be seen.) Open symbols are extrapolated values (dashed lines in Fig. 2d and Fig. 3d). Data on Nd-LSCO at $p = 0.07$, 0.12, 0.27, 0.36 and 0.40 (purple) are $\gamma$ values obtained on polycrystalline samples, as described in Fig. S9. Error bars are explained in the Supplementary Materials. We also include $\gamma$ for non-superconducting LSCO from published work (diamonds), obtained by extrapolating $C/T = \gamma + \beta T^2$ to $T = 0$ from data below 10 K ($p < 0.06$ (ref. 36); $p = 0.33$ (ref. 16)). The vertical dashed line marks the pseudogap critical point $p*$ in Nd-LSCO (Fig. 1). All solid lines are a guide to the eye.

[1] Cooper, R. A. *et al.* Anomalous criticality in the electrical resistivity of $La_{2-x}Sr_xCuO_4$. *Science* **323**, 603-607 (2009).

[2] Broun, D. M. What lies beneath the dome? *Nat. Phys.* **4**, 170-172 (2008).

[3] Collignon, C. *et al*. Fermi-surface transformation across the pseudogap critical point of the cuprate superconductor $La_{1.6-x}Nd_{0.4}Sr_xCuO_4$. *Phys. Rev. B*. **95**, 224517 (2017).

[4] Löhneysen, H. v. *et al.* Fermi-liquid instabilities at magnetic quantum phase transitions. *Rev. Mod. Phys.* **79**, 1015 (2007).

[5] Monthoux, P., Pines, D. & Lonzarich, G. G. Superconductivity without phonons. *Phys. Rev. Lett.* **450**, 1177-1183 (2007).

[6] Shibauchi, T., Carrington, A. & Matsuda, Y. A quantum critical point lying beneath the superconducting dome in iron pnictides. *Annu. Rev. Condens. Matter Phys.* **5**, 113-135 (2014).

[7] Löhneysen, H. v. *et al.* Non-Fermi-liquid behavior in a heavy-fermion alloy at a magnetic instability. *Phys. Rev. Lett.* **72**, 3262-3265 (1994).

[8] Walmsley, P. *et al.* Quasiparticle mass enhancement close to the quantum critical point in $BaFe_2(As_{1-x}P_x)_2$ . *Phys. Rev. Lett.* **110**, 257002 (2013).

[9] Doiron-Leyraud, N. *et al*. Correlation between linear resistivity and $T_c$ in the Bechgaard salts and the pnictide superconductor $Ba(Fe_{1-x}Co_x)_2As_2$. *Phys. Rev. B*. **80**, 214531 (2009).

[10] Daou, R. *et al.* Linear temperature dependence of resistivity and change in the Fermi surface at the pseudogap critical point of a high-$T_c$ superconductor. *Nat. Phys*. **5**, 31-34 (2009).

[11] Loram, J. W. *et al.* Electronic specific heat of $YBa_2Cu_3O_{6+x}$ from 1.8 to 300 K. *Phys. Rev. Lett.* **71**, 1740-1743 (1993).

[12] Loram, J. W. *et al.* Specific heat evidence on the normal state pseudogap. *J. Phys. Chem. Solids* **59**, 2091-2094 (1998).

[13] Ramshaw, B. J. *et al.* Quasiparticle mass enhancement approaching optimal doping in a high-$T_c$ superconductor. *Science* **348**, 317-320 (2015).

[14] Badoux, S. *et al.* Change of carrier density at the pseudogap critical point of a cuprate superconductor. *Nature* **531**, 201-214 (2016).

[15] Cyr-Choinière, O. *et al.* Pseudogap temperature *T\** of cuprate superconductors from the Nernst effect. *Phys. Rev. B* **97**, 064502 (2018).

[16] Nakamae, S. *et al.* Electronic ground state of heavily overdoped nonsuperconducting $La_{2-x}Sr_xCuO_4$. *Phys. Rev. B.* **68**, 100502R (2003).

[17] Matt, Ch. *et al.* Electron scattering, charge order, and pseudogap physics in $La_{1.6-x}Nd_{0.4}Sr_xCuO_4$: An angle-resolved photoemission spectroscopy study. *Phys. Rev. B* **92**, 134524 (2015).

[18] Verret, S. *et al.* Phenomenological theories of the low-temperature pseudogap: Hall number, specific heat, and Seebeck coefficient. *Phys. Rev. B* **96**, 125139 (2017).

[19] Trovarelli, O. *et al.* $YbRh_2Si_2$: pronounced non-Fermi-liquid effects above a low-lying magnetic phase transition. *Phys. Rev. Lett.* **85**, 626-629 (2000).

[20] Bianchi, A. *et al.* Avoided antiferromagnetic order and quantum critical point in $CeCoIn_5$. *Phys. Rev. Lett.* **91**, 257001 (2003).

[21] Momono, N. *et al.* Low-temperature electronic specific heat of $La_{2-x}Sr_xCuO_4$ and $La_{2-x}Sr_xCu_{1-y}Zn_yO_4$. Evidence for a *d*-wave superconductor. *Physica C* **233**, 395-401 (1994).

[22] Wade, J. M. *et al.* Electronic specific heat of $Tl_2Ba_2CuO_{6+\delta}$ from 2 K to 300 K for $0 < \delta < 0.1$. *J. Supercon.* **7**, 261-264 (1994).

[23] Bangura, A. F. *et al.* Fermi surface and electronic homogeneity of the overdoped cuprate superconductor $Tl_2Ba_2CuO_{6+\delta}$ as revealed by quantum oscillations. *Phys. Rev. B* **82**, 140501 (2010).

[24] Chan, M. K. *et al.* Single reconstructed Fermi pocket in an underdoped single-layer cuprate superconductor. *Nat. Commun.* **7**, 12244 (2016).

[25] Grissonnanche, G. *et al.* Direct measurement of the upper critical field in cuprate superconductors. *Nat. Commun.* **5**, 3280 (2014).

[26] Paul, I. & Kotliar, G. Thermoelectric behavior near the magnetic quantum critical point. *Phys. Rev. B* **64**, 184414 (2001).

[27] E. M. Motoyama *et al.* Spin correlations in the electron-doped high-transition-temperature superconductor $Nd_{2-x}Ce_xCuO_{4\pm\delta}$ . *Nature* **445**, 186 (2007).

[28] Dagan, Y. *et al.* Evidence for a quantum phase transition in $Pr_{2-x}Ce_xCuO_{4-\delta}$ from transport measurements. *Phys. Rev. Lett.* **92**, 167001 (2004).

[29] Fournier, P. *et al.* Insulator-metal crossover near optimal doping in $Pr_{2-x}Ce_xCuO_4$ : Anomalous normal-state low temperature resistivity. *Phys. Rev. Lett.* **81**, 4720-4723 (1998).

[30] Storey, J. *et al.* Pseudogap ground state in high-temperature superconductors. *Phys. Rev. B* **78**, 140506 (2008).

[31] Tranquada, J. M. *et al.* Coexistence of, and competition between, superconductivity and charge-stripe order in $La_{1.62-x}Nd_{0.4}Sr_xCuO_4$. *Phys. Rev. Lett.* **78**, 338 (1997).

[32] Nachumi, B. *et al.* Muon spin relaxation study of the stripe phase order in La$_{1.6-x}$Nd$_{0.4}$Sr$_x$CuO$_4$ and related 214 cuprates. *Phys. Rev. Lett.* **58**, 8760 (1998).

[33] Chatterjee, S. & Sachdev, S. Insulators and metals with topological order and discrete symmetry breaking. *Phys. Rev. B* **95**, 205133 (2017).

[34] Nie, L., Tarjus, G. & Kivelson, S. A. Quenched disorder and vestigial nematicity in the pseudogap regime of the cuprates. *PNAS* **111**, 7980-7985 (2014).

[35] Varma, C. M. Quantum-critical fluctuations in 2D metals: strange metals and superconductivity in antiferromagnets and in cuprates. *Rep. Prog. Phys.* **79**, 082501 (2016).

[36] Komiya, S. & Tsukada, S. Doping evolution of the electronic specific heat coefficient in slightly-doped La$_{2-x}$Sr$_x$CuO$_4$ single crystals. *J. Phys. Conference Series* **150**, 052118 (2009).

SUPPLEMENTARY INFORMATION

**Thermodynamic signatures of quantum criticality in cuprates**

B. Michon *et al.*

**CONTENTS**

**Methods**

**Supplementary Text**

# Methods

**Eu-LSCO**. Single crystals of $La_{2-y-x}Eu_ySr_xCuO_4$ (Eu-LSCO) were grown at the University of Tokyo with a Eu content $y = 0.2$, using a travelling-float-zone technique. Five samples were cut in the shape of small rectangular platelets, of typical dimensions 1 mm × 1 mm × 0.5 mm and mass of ~ 1 mg, from boules with nominal Sr concentrations $x = 0.08, 0.11, 0.16, 0.21$, and 0.24. The hole concentration $p$ is given by $p = x$, with a maximal error bar of ± 0.005. The critical temperature $T_c$ of all samples, defined as the onset of the drop in the magnetization, is plotted in Fig. S2a.

**Nd-LSCO**. Single crystals of $La_{2-y-x}Nd_ySr_xCuO_4$ (Nd-LSCO) were grown at the University of Texas at Austin with a Nd content $y = 0.4$, using a travelling-float-zone technique. Seven samples were cut in the shape of small rectangular platelets, of typical dimensions 1 mm × 1 mm × 0.5 mm and mass of ~ 1 mg, from boules with nominal Sr concentrations $x = 0.12, 0.15, 0.20, 0.22,$ 0.23, 0.24 and 0.25. The hole concentration $p$ is given by $p = x$, with an error bar ± 0.003, except for our sample with $p = 0.24$, for which the error bar is ± 0.005 [3]. The $T_c$ of all samples is plotted in Fig. S2b. The rapid decrease in $T_c$ from $p = 0.22$ to $p = 0.25$ provides a sensitive measure of the relative doping of samples in that range.

Powder samples of Nd-LSCO were produced at McMaster University with $y = 0.4$ and $x = 0.07$, 0.12, 0.27, 0.36 and 0.40, and shaped into sintered pellets. Samples with $x = 0.27, 0.36$ and 0.40 are not superconducting, and so their normal-state specific heat can be measured in zero field. The uncertainty on $p$ from $x$ is ± 0.01.

To perform the series of specific heat measurements reported in this article, we implemented an original AC modulation method which leads to an absolute accuracy of $\Delta C / C \sim \pm 4$ % with a relative sensitivity $\Delta C / C \sim \pm 0.01$ % on samples whose masses are on the order of 1.0 mg (down to 0.1 mg). These figures are valid for the temperature range 0.5 K < $T$ < 20 K and for the magnetic field range 0 < $H$ < 18 T.

**Calorimetric setup**. The calorimetric chips were prepared out of bare Cernox chips (1010 for $T < 2$ K, 1050 for $T > 2$ K). First, a shallow groove is made in the central part with a wire saw to obtain two independent films; one used as a heater (of resistance $R_h$) and the other one as the thermometer (of resistance $R_t$) (Fig. S12). The split chip is thereafter attached to a small copper ring with PtW(7%) wires, 25 or 50 μm in diameter and 1 to 2 mm in length, glued with a minute amount of Ag epoxy. The choice of wires is important since it defines the external thermal conductance and the frequency range where the measurements will have their optimal accuracy (between 0.5 Hz and 20 Hz). We used PtW wires since their thermal conductivity is insensitive to magnetic fields (< 1 % in 18 T).

**Modulation technique**. In the simplest approximation, when an alternative current $I_{ac}$ is applied at a frequency $\omega$ on the heater side, temperature oscillations $T_{ac} = P_{ac} / (K_e + jC2\omega)$ are induced at $2\omega$, where $K_e$ is the external thermal conductance and $C$ the total heat capacity. Introducing the phase $\varphi$ of $T_{ac}$ relative to the power $P_{ac}$, one gets:

$C = P_{ac} \sin(-\varphi) / [2\omega |T_{ac}|]$ , with $P_{ac} = R_h |I_{ac}|^2 / 2$ .

These oscillations can then be measured by applying a DC current $I_{dc}$ across the thermometer: $V_{ac}^t(2\omega) = [(dR_t / dT) \, T_{ac}(2\omega)] \, I_{dc}$ . The main drawback of this simple approach is that internal time corrections (due to finite thermal conductances between the different parts of the chips: thermometer-heater-substrate-sample) are not taken into account. To overcome this difficulty, especially at the lowest temperature where the thermal conductivity between the sensing layer of the Cernox and the sapphire substrate is the main limiting factor, we have also used the heater side to measure the temperature oscillations: $V_{ac}^h(3\omega) = [(dR_h / dT) \, T_{ac}(2\omega)] \, I_{ac}(\omega)$ .

When all the internal conductances are larger than the external one $K_e$, the $T_{ac}$ measured on the thermometer side (at $2\omega$) must be the same as that measured on the heater side (at $3\omega$). Any difference points to a thermal gradient within the chips, which can then be minimized by adjusting the measurement parameters. This procedure improves the absolute accuracy and enables a much better estimate of the error bars.

**Thermometry**. The first step is to know precisely the temperature of the main heat sink on which the measuring chip is attached. This is achieved with commercial calibrated Cernox sensors 1010 and 1050, used respectively in the ranges 0.5 − 5 K and 1.5 − 20 K. The calibration has been further improved in-house with a superconductive fixed point device and a CMN thermometer to reach an accuracy of ± 1 % in the absolute temperature, within the temperature range considered here. All thermometers have then been thoroughly calibrated in field, from 0.2 K to 4 K against a Ge sensor placed in a compensated area of a 18 tesla superconducting magnet and between 2 K and 20 K against a capacitor. The output of this protocol is a quintic bivariable ($T$ and $H$) spline interpolation sheet, which was used to determine and control the temperature between 0.5 and 20 K up to 18 T with a relative accuracy $\Delta T/T \sim \pm 0.2$ %.

**Addenda and test**. In order to subtract the heat capacity of the sample mount (chip + a few μg of Apiezon grease used to glue the sample onto the back of the chip), the empty chip (with grease) was measured prior to each sample measurement. This background heat capacity is on the order of $C_{add}/T \sim 5$ nJ / K$^2$ at 1 K (3 K) for 1010 (1050) chips, which represents 10 % to 50 % of the heat capacity of the samples.

To benchmark our measurement system and technique, we measured a piece of ultrapure copper, of mass 1 mg, whose heat capacity at low $T$ is comparable to that of our Eu-LSCO and Nd-LSCO samples. In zero magnetic field, the reproducibility between different runs and between different chips shows an absolute accuracy of at least $\Delta C/C \sim \pm 3$ % compared to NBS tabulated values, across the range from 0.5 K to 10 K (Fig. S13). In magnetic fields up to 18 T, the apparent change $\Delta C / C$ vs $H$ is very smooth and does not exceed 1%.

All factors considered, the absolute accuracy of each of the specific heat runs is estimated to be $\Delta C / C \sim \pm 4\%$, or better. For 0.5 K < $T$ < 5 K, the specific heat is dominated by the electronic contribution and $\Delta C_{el}/C_{el} \sim \Delta C / C$ at $T = 0.5$ K and 1.0 K. But this error bar increases with temperature, due to the rapid increase in the phonon contribution (and the Schottky term in Nd-LSCO), which makes the electronic term a smaller and smaller fraction of the total signal. The error bars plotted in Fig. 4b show how this translates into an uncertainty on the absolute value of $C_{el}/T$ for each sample separately, at each temperature.

NEUTRON POWDER MEASUREMENTS

Inelastic neutron scattering measurements were performed on 12 samples of La$_{1.6-x}$Nd$_{0.4}$Sr$_x$CuO$_4$ with 0.01 < $x$ < 0.4 using the direct geometry chopper instrument, SEQUOIA, BL-17 [37] at the Spallation Neutron Source, Oak Ridge National Laboratory. The ~ 10 gram powder samples were loaded into aluminum sample cans with helium gas to ensure good thermal equilibrium, and cooled in a closed-cycle refrigerator with a base temperature of 4 K. Measurements at $T = 4$ K, 35 K and 200 K were performed with incident energies of 60 meV and 11 meV on all samples. Inelastic measurements with $E_i$ = 60 meV employed a $T_0$ chopper at 90 Hz, and Fermi chopper 2 set to 420 Hz. The energy resolution of the $E_i$ = 60 meV data was ~ 2% of $E_i$ (or 1.2 meV), at the elastic position; the resolution improved for neutron-energy-loss inelastic scattering. The data were reduced using MANTID [38] and visualized and analyzed using DAVE [39] software package. The results are presented in Fig. S8. They show that the energy of the acoustic phonons in Nd-LSCO increases by less than 3 % from $x$ = 0.16 to $x$ = 0.24.

## Supplementary Text

EXTRACTING THE ELECTRONIC SPECIFIC HEAT

There are three contributions to the specific heat C of Eu-LSCO and Nd-LSCO: from electrons ($C_{el}$), phonons ($C_{ph}$) and nuclei ($C_{nuclear}$). In Nd-LSCO, there is an additional contribution from the magnetic moment on the $Nd^{3+}$ ions ($C_{mag}$). Because Eu ions are not magnetic, this term is absent in Eu-LSCO. To extract the electronic term of interest here, we proceed as outlined in Fig. S7, and described in detail below.

Nuclear Schottky term : $C_{nuclear}$

The nuclear hyperfine term is a Schottky anomaly peaked at very low temperature. Above the peak, $C_{nuclear}$ dies off rapidly, as $C_{nuclear} \sim 1 / T^2$. In Eu-LSCO and Nd-LSCO, $C_{nuclear}$ is clearly visible in all samples as a rapid upturn below 1 K (Fig. 2b, Fig. 3c, Fig. S4a, Fig. S4c). In a field of 8 T, $C_{nuclear}$ becomes negligible above $T \sim 1$ K (Fig. 2c, Fig. S7a); in 18 T, above $\sim 2$ K (Fig. 3a). By subtracting the raw data for C vs T in Eu-LSCO at $p = 0.16$ from the raw data $C(T)$ at each other doping, we see that the rapid upturn below 1 K is almost entirely removed (Fig. 2d, Fig. S4b, Fig. S7b), at least down to 0.5 K. (Below $T = 0.5$ K, we do observe that the subtraction is not perfect, reflecting a small difference in the nuclear Schottky anomaly from sample to sample (within $\pm$ 20% at 0.3 K), and this is why we report data only for $T = 0.5$ K and above.) This shows that $C_{nuclear}$ is very weakly sample-dependent and doping-dependent. (The nuclear Schottky anomaly is believed to come from the Eu ions [40], and the Eu content of all our Eu-LSCO samples is kept fixed.)

In summary, to remove $C_{nuclear}$ from the measured C we can work at $T = 1$ K (in 8 T) or $T = 2$ K in 18 T, and higher temperatures. Below 1 K, we can remove $C_{nuclear}$ reliably down to 0.5 K by subtracting a reference curve, for example $p = 0.16$ in Eu-LSCO and $p = 0.12$ in Nd-LSCO, since $C_{nuclear}$ varies from sample to sample by less than $\pm$ 5% at 0.5 K. In Fig. S7B, we see the effect of such a subtraction: $(C - C_{nuclear}) / T$ is constant below 1 K at $p = 0.11$ and $p = 0.16$, while it rises monotonically as $T \rightarrow 0$ at $p = 0.24$. (Note that 8 T is enough to suppress superconductivity in Eu-LSCO down to the lowest temperature for all dopings except $p = 0.21$, where we use 18 T and are limited to $T > 2$ K (Fig. 2d).)

Magnetic Schottky term : $C_{mag}$

Compared to Eu-LSCO, the specific heat of Nd-LSCO has one additional contribution, $C_{mag}$ , a Schottky peak due to the magnetic moment on the $Nd^{3+}$ ions (Fig. S6b). A field moves this magnetic Schottky peak up in temperature, so that $C_{mag}$ becomes very small below 2 K in 8 T and very small below 6 K in 18 T (Fig. 3a). We apply the same subtraction procedure as for

Eu-LSCO, using our Nd-LSCO sample with $p = 0.12$ as the reference. (The subtraction procedure works well to remove both $C_{mag}$ and $C_{nuclear}$ because the Nd content of all our Nd-LSCO samples is kept fixed.)

At high doping ($p > 0.25$), single crystals are difficult to grow and we have therefore resorted to powder samples of Nd-LSCO, with $p = 0.27$, $0.36$ and $0.40$. Because these samples are not superconducting, the normal state specific heat can be measured in zero field. Note, however, that the magnetic Schottky term depends on the field direction relative to the $c$ axis of the crystal structure. As a result, we could not use a field to suppress $C_{mag}$ in the powder samples, made of micro-crystallites of all orientations. In Fig. S9a, we compare directly the raw data from our powder sample at $p = 0.12$ with the raw data from our single-crystal sample at $p = 0.12$, both at $H = 0$. The two sets of raw data are seen to be in excellent agreement. In the single crystal, we remove $C_{mag}$ by applying a field of 18 T; a fit to $C / T = \gamma + \beta T^2$ below ~ 5 K yields $\gamma = 3.6$ mJ / $K^2$ mol and $\beta = 0.215$ mJ / $K^4$ mol (Fig. 3b). In the powder sample, we fit the zero-field raw data to a Schottky anomaly that goes as $C_{mag} \sim 1 / T^2$ (in the range 3 K < T < 7 K). We subtract $C_{mag}$ from the raw powder data and fit the difference to $(C - C_{mag}) / T = \gamma + \beta T^2$ (Fig. S9b). The resulting values of $\gamma$ ($4 \pm 1$ mJ / $K^2$ mol) and $\beta$ ($0.225 \pm 0.015$ mJ / $K^4$ mol) are in good agreement with those quoted above for the $p = 0.12$ single crystal.

Performing the same fit and subtraction to the zero-field raw powder data at $p = 0.27$, $0.36$ and $0.40$ yields curves that are roughly parallel when plotted as $(C - C_{mag}) / T$ vs $T^2$, shifted rigidly upwards relative to the $p = 0.12$ curve (Fig. S9b). A fit to $(C - C_{mag}) / T = \gamma + \beta T^2$ yields the values of $\gamma$ plotted in Fig. 4b (purple dots) and the values of $\beta$ plotted in Fig. S7d (orange dots).

Phonon term : $C_{ph}$

At 18 T, superconductivity is completely eliminated from all samples (Fig. S6b). At $T > 2$ K, $C_{nuclear}$ is negligible. In Nd-LSCO, $C_{mag}$ is negligible below ~ 6 K. So in the range 2 – 6 K at 18 T, we have $C = C_{el} + C_{ph}$. In Eu-LSCO at $p = 0.11$ and $0.16$, and in Nd-LSCO at $p = 0.12$, the raw data in that $T$ range are well described by $C / T = \gamma + \beta T^2$ (Fig. 2a, Fig. 3b), with very similar values of $\beta$ (Fig. S7d). Alternatively, we can fit the 8 T data from those two samples of Eu-LSCO over the full range from $T = 0.5$ K to 10 K by using $(C - C_{nuclear}) / T = \gamma + C_{ph} / T$, where $C_{ph} / T = \beta T^2 + \delta T^4$, as shown in Fig. S7. The two approaches yield very similar values of $\gamma$ and $\beta$ (identical within error bars). Fitting single-crystal data away from $p^* = 0.23$, namely Eu-LSCO at $p = 0.08$, $0.11$, and $0.16$ in 8 T and Nd-LSCO at $p = 0.12$, $0.15$, and $0.20$ in 18 T, to $(C - C_{nuclear}) / T = \gamma + \beta T^2 + \delta T^4$ yields the values of $\beta$ plotted in Fig. S7d. For the four powder samples of Nd-LSCO, we fit the zero-field data to $(C - C_{mag}) / T = \gamma + \beta T^2 + \delta T^4$. In Fig. S7D, we see that the doping dependence of $\beta$ is weak, and in excellent agreement with the doping dependence of the phonon energy, $E_{ph}$, measured by neutron scattering (Fig. S8), *i.e.* $\beta \sim 1 / E_{ph}^3$. The neutron data predict a decrease in $\beta$ of $\sim 7\%$ between $p = 0.16$ and $p = 0.24$ and $\sim 10\%$ between $p = 0.12$ and $p = 0.24$.

Doping dependence of the electronic term : $C_{el}$ vs $p$

In Fig. S7a, we see that $C_{nuclear}$ in 8 T is negligible at $T = 1$ K (and above). In Fig. S7b, we see that $C_{ph}$ is negligible at $T = 1$ K (and below). Therefore the raw data at $T = 1$ K and $H = 8$ T, in Eu-LSCO, directly give the electronic specific heat at 1 K, *i.e.* $C_{el} = C$ to a very good approximation (except at $p = 0.21$, where 8 T is not enough to fully suppress superconductivity). The huge rise in $C$ from $p = 0.11$ to $p = 0.24$ (Fig. 4a) is therefore entirely due to $C_{el}$.

At high doping, beyond $p^*$, the Nd-LSCO powder data show a clear decrease in $\gamma$ as $p$ increases (Fig. S9b). The result is therefore an unambiguous peak, at $p = p^*$, in $C_{el}$ vs $p$ in the $T = 0$ limit (Fig. 4a). This is the first key signature of a QCP.

Temperature dependence of the electronic term : $C_{el}$ vs $T$

Fig. S7 shows that $C_{el} / T$ at $p = 0.16$ is constant as a function of $T$, all $T$ dependence being due to $C_{nuclear} / T$ and $C_{ph} / T$. The same is true at $p = 0.11$. At $p = 0.24$, however, there is an additional $T$ dependence not due to $C_{nuclear}$ or $C_{ph}$, which comes from $C_{el}(T)$. The simplest way to reveal this $T$ dependence of $C_{el}(T)$ is to subtract the raw curve at $p = 0.16$ from the raw curve at $p = 0.24$, as done in Fig. 2d. Or, equivalently, subtract the fit to $(C - C_{nuclear} - C_{ph})/T$ performed on the $p = 0.16$ data, as done in Fig. S7c. Both approaches yield a clean $\log(1/T)$ dependence for $C_{el} / T$ at $p = 0.24$, from 0.5 K to 10 K (see dotted line in Fig. S7c).

These approaches assume that $C_{nuclear}$ is the same in the two samples, which is true to better than 5%, and also that $C_{ph}$ at $p = 0.24$ is the same as $C_{ph}$ at $p = 0.16$, which we know is not quite true. Indeed, our neutron data indicate that $C_{ph}$ is smaller at $p = 0.24$ by $\sim 7\%$. In Fig. S7c, we take this correction into account, multiplying $C_{ph}$ by a factor 0.93 before subtracting it from $(C - C_{nuclear})$ to obtain $C_{el}(T)$ at $p = 0.24$. The resulting grey curve shows that this 7% correction has little effect below 5 K, and the small modification it makes between 5 K and 10 K is within the uncertainty of the measurement (Fig. S7c). In other words, within error bars, we still find that $C_{el} / T \sim \log(1/T)$ up to at least 10 K. This is the second key signature of a QCP.

In Fig. 4b, we plot the value of $C_{el} / T$ at $T = 0.5$ K, 2 K and 10 K, obtained using the first procedure of simply subtracting two raw curves, requiring no fitting at all. As we have just shown, the values of $C_{el} / T$ obtained in this way are accurate and reliable, within the error bars.

In summary, the high level of quantitative consistency we find between the values of $C_{el} / T$

obtained in our 5 crystals of Eu-LSCO and those obtained in our 7 crystals of Nd-LSCO (Fig. 4b) is a strong validation of both the experimental technique and the data analysis. It confirms that the electronic specific heat $C_{el}(T)$ we report is reproducible and accurate (within the quoted error bars). Furthermore, the excellent agreement between the values of $C_{el}(T)$ plotted in Fig. 4b and the raw values of $C(T)$ at $T = 1$ K and 2 K plotted in Fig. 4a confirms the fidelity of our analysis.

VAN HOVE SINGULARITY

In hole-doped cuprates, the large Fermi surface in the overdoped regime undergoes a change of topology from hole-like to electron-like at some material-dependent critical doping $p_{vHs}$ . According to ARPES, this change of topology occurs between $p = 0.15$ and $p = 0.22$ in LSCO [41] and between $p = 0.20$ and $p = 0.24$ in Nd-LSCO [17], so close to $p^*$ in both cases. If the Fermi surface were strictly two-dimensional, this would, in the clean limit, lead to a van Hove singularity in the density of states, producing a cusp in $C_{el}$ / $T$ vs $p$ at $p_{vHs}$ and a $\log(1/T)$ dependence of $C_{el}$ / $T$ as $T \to 0$, analogous to the behavior we report in Eu-LSCO and Nd-LSCO.

However, the change of Fermi-surface topology at $p_{vHs} \sim p^*$ cannot in fact be responsible for that behavior, for two reasons: because of the significant 3D character of the Fermi surface in Nd-LSCO (and Eu-LSCO), and because of the significant level of disorder in our samples. Each of these two mechanisms broadens the van Hove singularity, removes the cusp vs $p$ and cuts off the $\log(1/T)$ divergence at low $T$.

To be quantitative, we have calculated the specific heat of Nd-LSCO associated with its known band structure, using the following one-band model [42]:

$$\xi_{k} = -\mu - 2t(\cos(k_x) + \cos(k_y)) - 4t'\cos(k_x)\cos(k_y) - 2t''(\cos(2k_x) + \cos(2k_y))$$
$$- 2t_z\cos(k_z/2)\left[\cos(k_x) - \cos(k_y)\right]^2 \cos(k_x/2)\cos(k_y/2)$$

with $t = 0.189$ eV, $t'/t = -0.17$, $t''/t = 0.05$. These parameters agree with the experimental band structure measured by ARPES [43], and are such that the vHs is located at $p = 0.23$. The 3D character is controlled by the interlayer hopping parameter $t_z$. The doping is adjusted by tuning the chemical potential $\mu$.

The temperature-dependent specific heat is given by [18] :

$$C_{el}(T) = \int_{-\infty}^{\infty} d\epsilon \, \frac{df(\epsilon)}{dT} N(\epsilon),$$

with:

$$N(\omega) = -\sum_{\mathbf{k}} \frac{1}{\pi} \frac{\hbar/2\tau}{(\omega - \xi_{\mathbf{k}})^2 + (\hbar/2\tau)^2}.$$

where $f(\xi)$ is the Fermi-Dirac function, $N(\xi)$ is the density of state, and $\tau$ is the quasiparticle lifetime.

The calculated specific heat is shown in Fig. S10 (blue), as a function of doping (at $T \to 0$) on the left and as a function of temperature (at $p = p_{vHs}$) on the right. In the top panels, we show the clean-limit 2D result, with $\hbar / \tau = 0$ and $t_z = 0$. In the middle panels, we show the effect of disorder, with $\hbar / \tau = t / 25$, the value needed to produce the measured residual resistivity of our Nd-LSCO and Eu-LSCO samples at $p = 0.24$, namely $\rho_0 \sim 30$ μΩ cm (Fig. S3). In the lower panels, we further add the effect of 3D dispersion, with $t_z \equiv 0.13t$, the value needed to produce the measured anisotropy of Nd-LSCO at $p = 0.24$, namely $\rho_c / \rho_a \sim 250$ [10].

Direct comparison with our data shows that band structure effects associated with $p_{vHs}$ in Nd-LSCO do produce a broad background bump in $C_{el}$ vs $p$, but they cannot account for the large and sharp peak we observe in $C_{el} / T$ at $T = 0.5$ K (Fig. S10). Moreover, the fact that we see $C_{el} / T \sim \log(1/T)$ persisting down to 0.5 K at $p = p^*$ completely excludes a van Hove mechanism, which yields a constant $C_{el} / T$ when $k_B T < \hbar / \tau$ or when $k_B T < t_z$, *i.e.* below $\sim 20$ K in our samples (Fig. S10).

## Supplementary References

[37] G. E. Granroth *et al.*, J. Phys. Conf. Ser. **251**, 012058 (2010).

[38] O. Arnold *et al.*, Nuclear Instruments and Methods in Physics Research Section A: Accelerators, Spectrometers, Detectors and Associated Equipment **764**, 156 (2014).

[39] R. Azuah *et al.*, J. Res. Natl. Inst. Stan. Technol. **114**, ppp (2009).

[40] S. Ghamaty *et al.*, *Physica C* **160**, 217 (1989).

[41] T. Yoshida *et al.*, *Phys. Rev. B* **74**, 224510 (2006).

[42] R. S. Markiewicz *et al.*, *Phys. Rev. B* **72**, 054519 (2005).

[43] C. E. Matt *et al.*, *arXiv:1707.08491* (2017).

[44] C. Marcenat *et al.*, *Nat. Commun.* **6**, 7927 (2015).

[45] J. W. Loram *et al.*, *Physica C* **282-287**, 1405 (1997).

[46] C. Collignon *et al.*, unpublished.

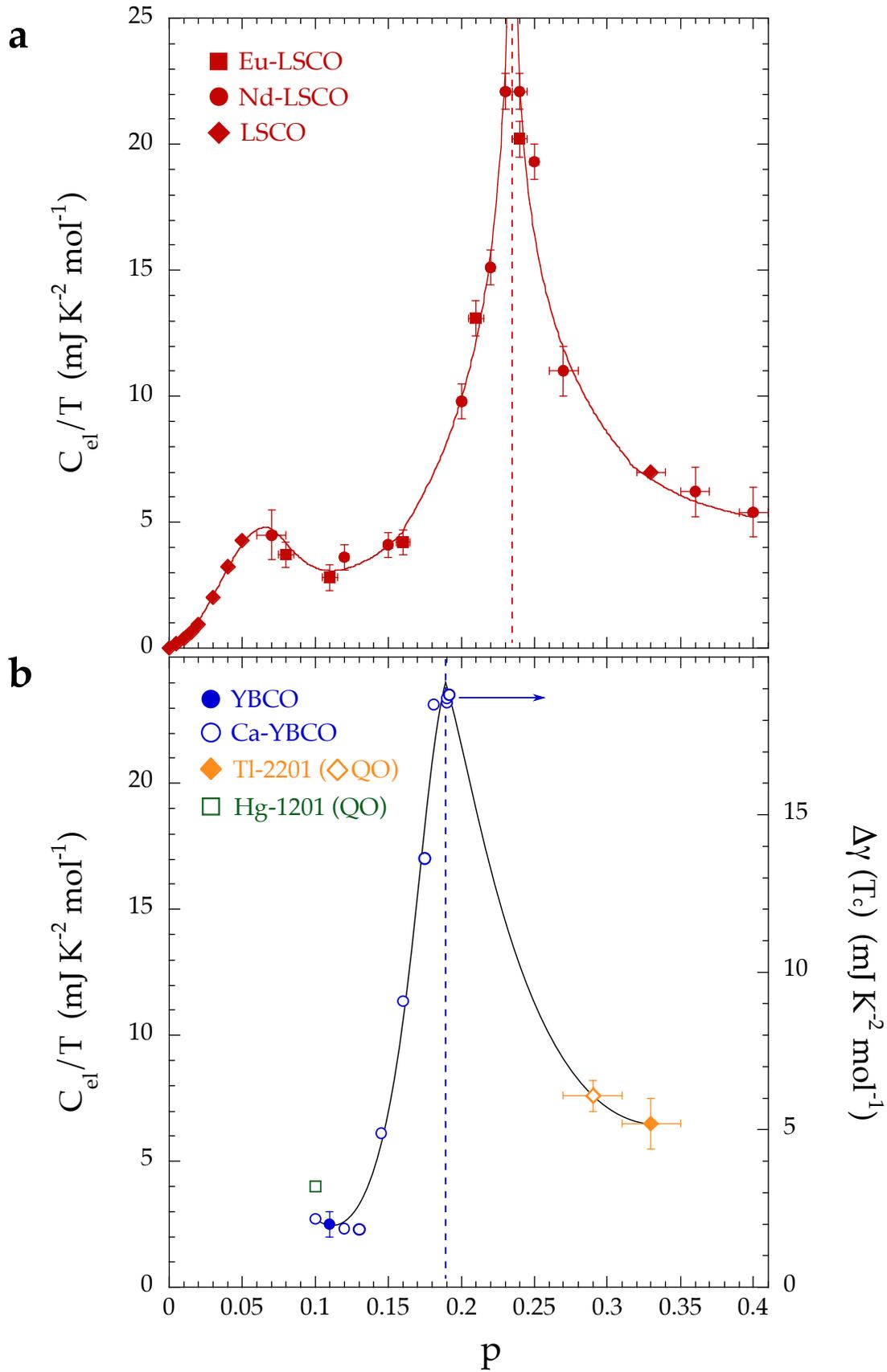

**Fig. S1 | Comparison with data in Tl-2201, YBCO and Hg-1201.**

**a)** Normal-state electronic specific heat of Eu-LSCO (squares), Nd-LSCO (circles), and LSCO (diamonds) at $T = 0.5$ K (from Fig. 4b). **b)** Electronic specific heat coefficient $\gamma$ in non-superconducting Tl-2201 at $p = 0.33 \pm 0.02$ (full orange diamond, left axis; ref. 22) and in the field-induced normal state of YBCO at $p = 0.11$ (full blue circle, left axis; ref. 44). Also shown are the values of $\gamma$ obtained from the effective mass $m^*$ measured by quantum oscillations in Tl-2201 at $p = 0.29 \pm 0.02$ (open orange diamond, left axis; ref. 23) and in Hg-1201 at $p \sim 0.1$ (open green square, left axis; ref. 24). The jump in specific heat at $T_c$ measured in Ca-doped YBCO (open blue circles; ref. 45) is plotted for comparison (right axis). The solid lines are a guide to the eye. We observe a strong quantitative similarity between the LSCO-based materials on the one hand (panel a) and YBCO–Hg-1201–Tl-2201 on the other (panel b), apart from a shift in the peak position (dashed lines) that tracks $p^*$ (0.19 in YBCO, 0.23 in Nd-LSCO).

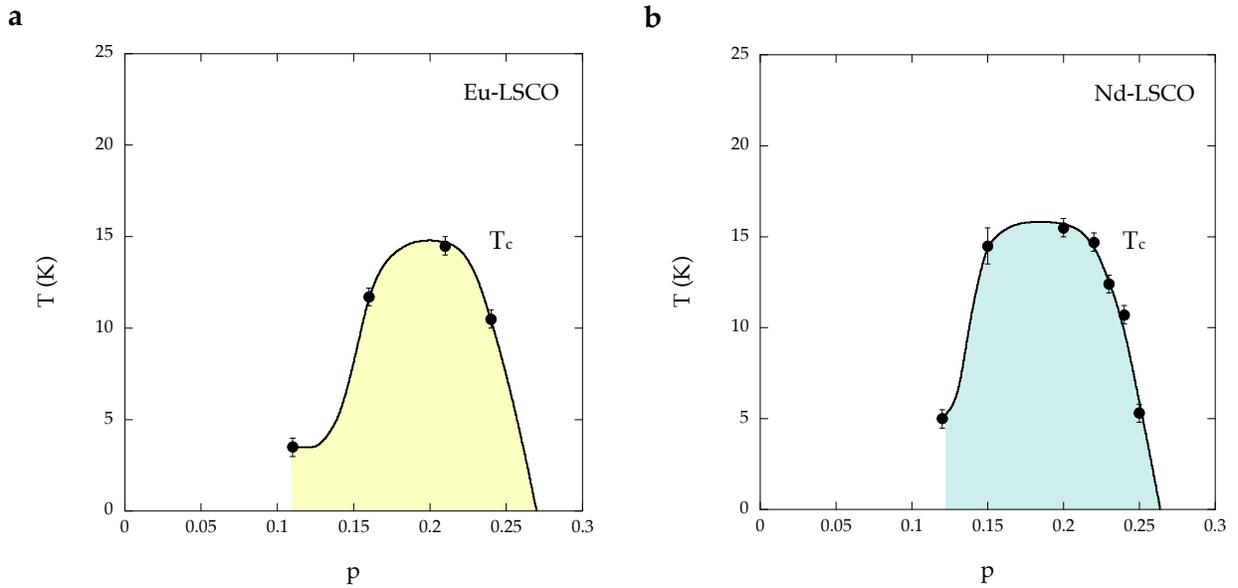

**Fig. S2 | $T_c$ vs doping in our samples of Eu-LSCO and Nd-LSCO.**

**a)** $T_c$ vs $p$ in Eu-LSCO. **b)** $T_c$ vs $p$ in Nd-LSCO. $T_c$ is defined as the onset of the drop in the magnetization upon cooling.

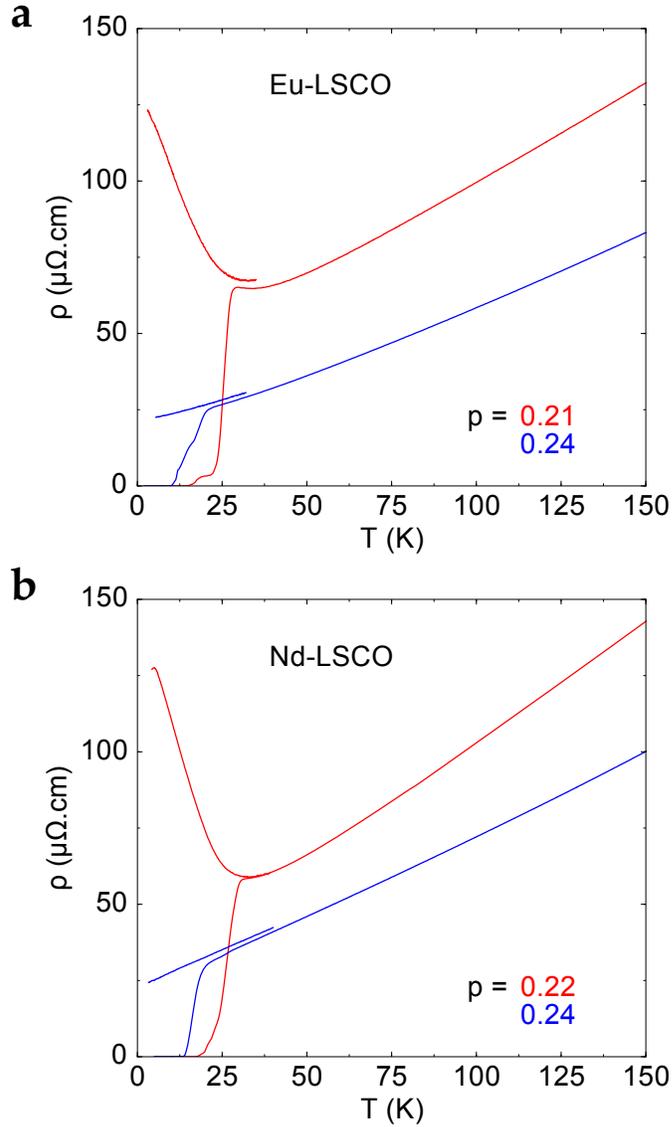

**Fig. S3 | Temperature dependence of the resistivity in Eu-LSCO and Nd-LSCO.**

**a)** $\rho$ vs $T$ in our Eu-LSCO samples with $p = 0.21$ (red) and $p = 0.24$ (blue), at $H = 0$ and $H = 33$ T (short section below 40 K) [46]. **b)** Same for our Nd-LSCO samples with $p = 0.22$ (red) and $p = 0.24$ (blue) [3]. The approximately linear $\rho(T)$ as $T \to 0$ at $p = 0.24$ (blue) shows that 0.24 is close to the critical point $p^* \sim 0.23$ in both materials. The large upturn in $\rho(T)$ as $T \to 0$ at $p = 0.21$ and $p = 0.22$ (red) shows that the pseudogap has opened in both materials (at $p < 0.23$).

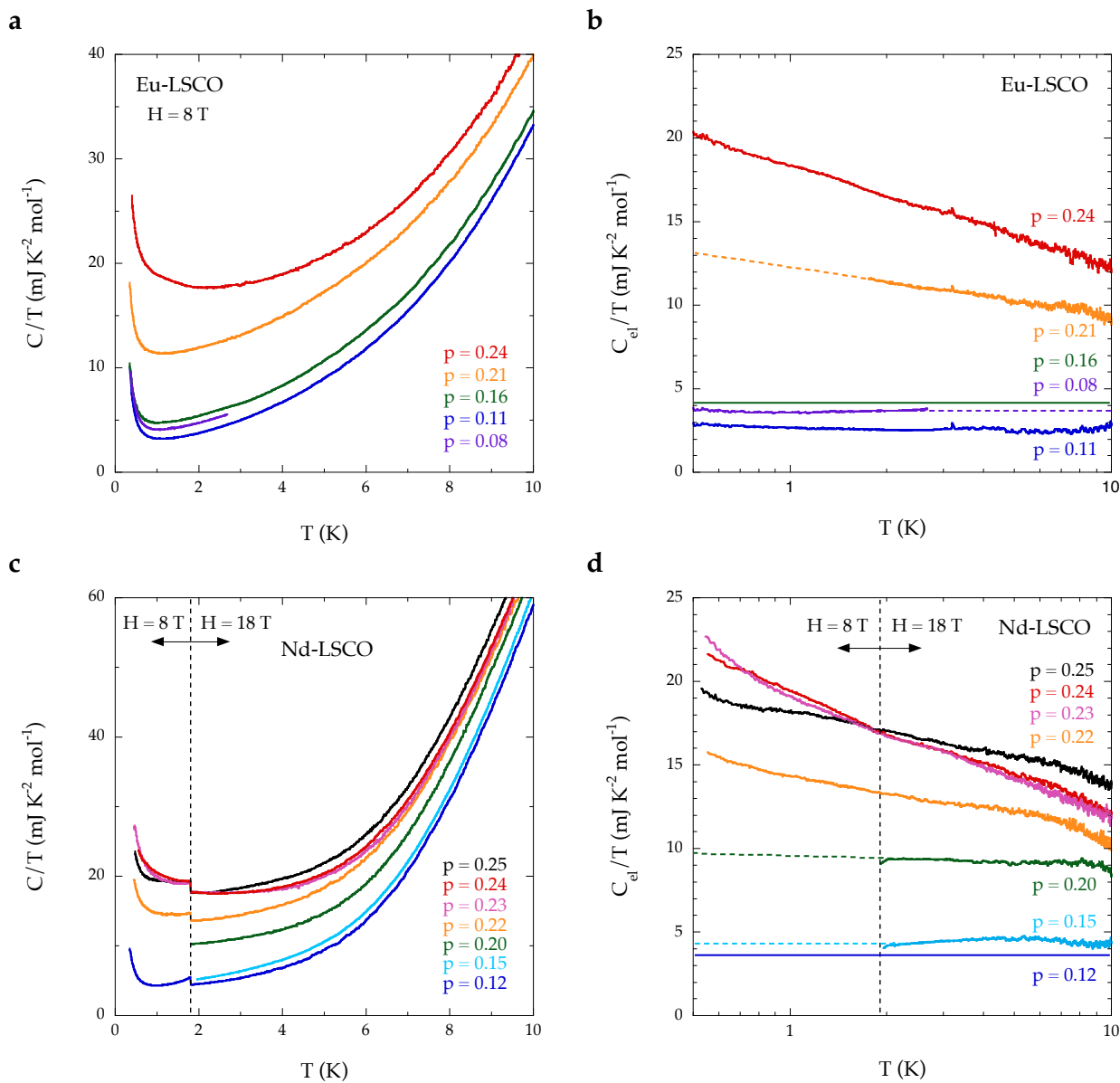

**Fig. S4 | Specific heat data for all crystals of Eu-LSCO and Nd-LSCO.**

**b)** Same as Fig. 2b, for all 5 Eu-LSCO crystals. **b)** Same as Fig. 2d, for those 5 crystals.
**c)** Same as Fig. 3c, for all 7 Nd-LSCO crystals. **d)** Same as Fig. 3d, for those 7 crystals.

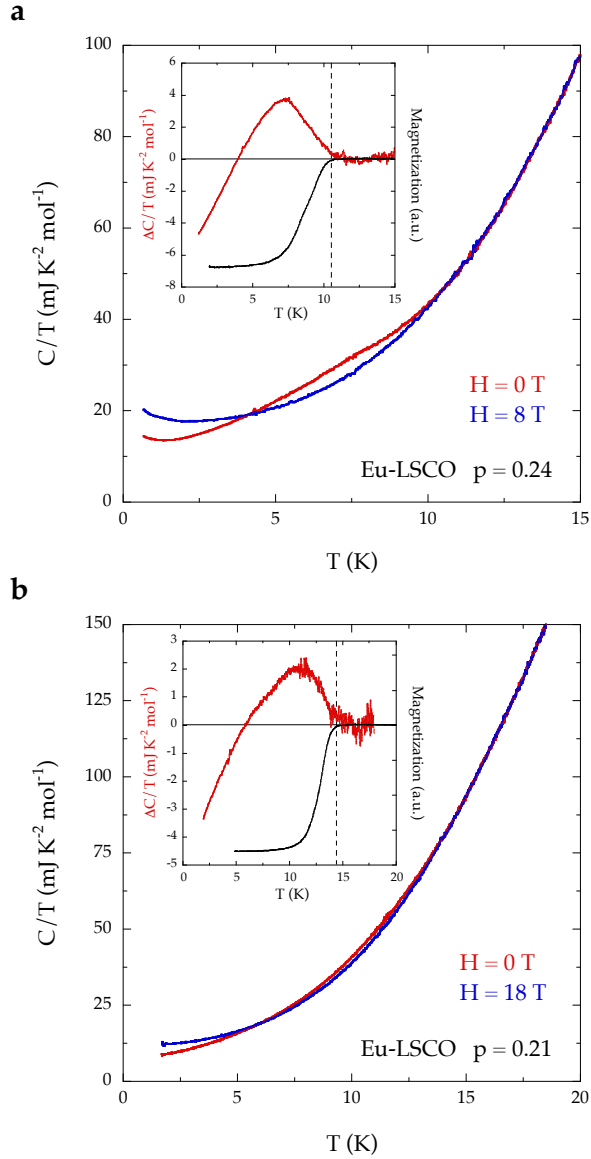

**Fig. S5 | Specific heat of Eu-LSCO as a function of temperature.**

**a)** $C / T$ vs $T$ in our Eu-LSCO sample with $p$ = 0.24, at $H$ = 0 (red) and $H$ = 8 T (blue). *Inset*: Difference between the two curves in the main panel (red). This is the difference between the superconducting-state $C / T$ and the normal-state $C / T$. The black curve is the magnetization of that sample. At $p$ = 0.24, the bulk $T_c$ = 10.5 ± 0.5 K (dashed line). **b)** As in panel a, for our sample with $p$ = 0.21, at $H$ = 0 (red) and $H$ = 18 T (blue). *Inset*: as in Panel a. At $p$ = 0.21, the bulk $T_c$ = 14.5 ± 0.5 K.

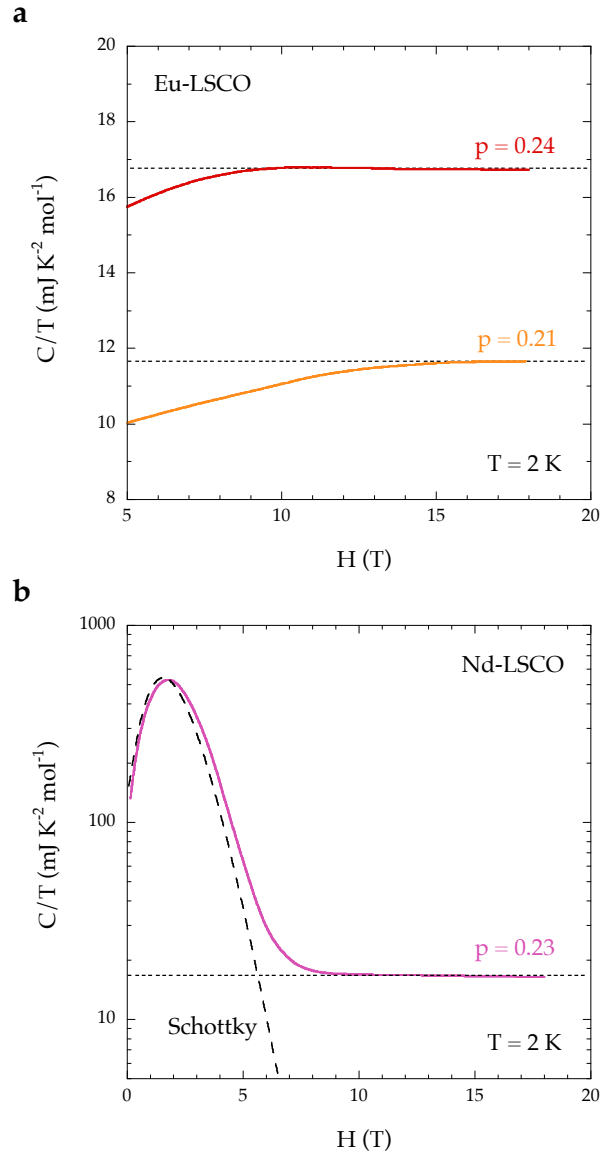

**Fig. S6 | Specific heat as a function of magnetic field.**

**a)** $C / T$ vs $H$ in our Eu-LSCO samples with $p = 0.21$ (orange) and $p = 0.24$ (red), at $T = 2$ K. The upper critical field above which there is no remaining superconductivity is $H_{c2} = 15$ T at $p = 0.21$ and $H_{c2} = 9$ T at $p = 0.24$. Note that for $p = 0.24$, $C / T$ has reached 99 % of its normal state value by 8 T. **b)** $C / T$ vs $H$ in our Nd-LSCO sample with $p = 0.23$, at $T = 2$ K, in a semi-log plot. The dashed line shows the expected field dependence of the Schottky contribution associated with Nd ions ($C_{mag}$). The data are independent of field above $H \sim 9$ T.

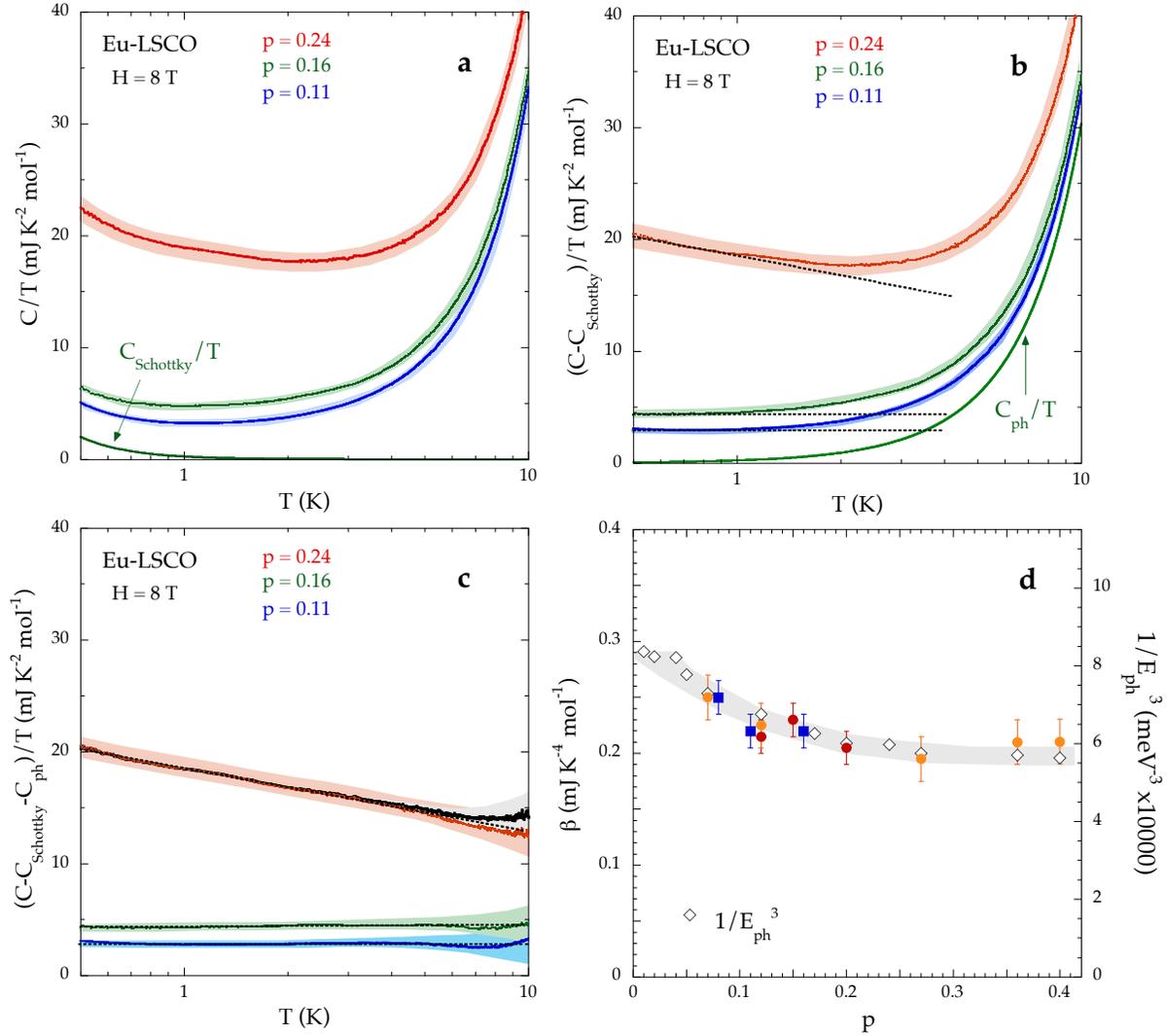

**Fig. S7 | Doping dependence of phonon specific heat in Eu-LSCO and Nd-LSCO.**

**a)** Raw data for Eu-LSCO at $p = 0.11$ (blue), 0.16 (green) and 0.24 (red), plotted as $C / T$ vs $\log T$. The width of the pale band tracking each curve is the uncertainty on the absolute measurement of $C$ ($\pm 4\%$). The dotted lines are explained in b). The solid green line is a fit to $C_{\text{nuclear}} \sim T^{-2}$ for the $p = 0.16$ data. **b)** Same three curves as in panel a, from which the same Schottky anomaly, $C_{\text{nuclear}} / T$, the green line in panel a, has been subtracted. The dotted lines show that $(C - C_{\text{nuclear}}) / T$ is flat as $T \rightarrow 0$ for $p = 0.11$ and 0.16, while it rises as $\log(1/T)$ for $p = 0.24$. The solid green line is a fit of the green curve at $p = 0.16$ to $(C - C_{\text{nuclear}}) / T = \gamma + C_{\text{ph}} / T$ up to 10 K, where $C_{\text{ph}} / T = \beta T^2 + \delta T^4$ is the phonon contribution. **c)** Same three curves as in panel b, from which the same phonon contribution $C_{ph} / T$, the green line in panel b, has been subtracted. We see that within error bars the resulting $C_{el} / T$ is constant up to 10 K for $p = 0.11$ (blue) and 0.16 (green), while it varies as $\log(1/T)$ up to 10 K for $p = 0.24$ (red). The grey curve is obtained by subtracting 0.93 $C_{ph}/T$ from the $p = 0.24$ data, instead of $C_{ph}/T$. This shows that the 7% decrease in $C_{ph}$ expected from our neutron data in going from $p = 0.16$ to $p = 0.24$ (see panel d) has no impact on the $\log T$ behavior below 5 K. **d)** Doping dependence of the phonon specific heat parameter β , in $C_{ph} / T = \beta T^2 + \delta T^4$, obtained from a fit to $(C - C_{nuclear}) / T = \gamma + C_{ph} / T$ up to 10 K, for Eu-LSCO crystals (dark blue squares), Nd-LSCO crystals (red dots), and Nd-LSCO powders (orange dots). The open diamonds are $1 / E_{ph}^3$ (right axis), where $E_{ph}(p)$ is the phonon energy measured by neutron scattering (Fig. S8). The grey band corresponds to a $\pm$ 1% uncertainty on $E_{ph}$. We see that $\beta \sim 1/ E_{ph}^3$ across the full doping range.

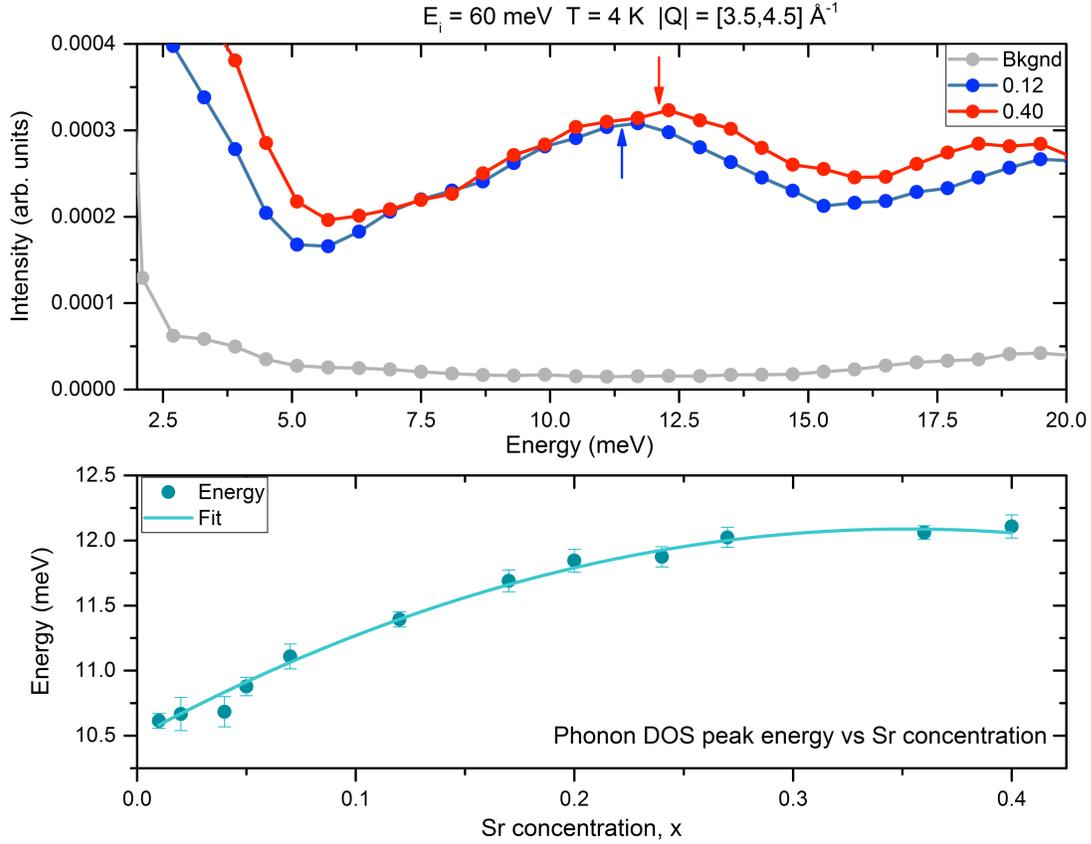

**Fig. S8 | Doping dependence of phonon density of states in Nd-LSCO.**

<u>Upper panel</u>: Neutron scattering intensity vs energy at $T = 4$ K, measured on powder samples for Nd-LSCO at $x = 0.12$ (blue) and $x = 0.40$ (red) at the SNS facility in Oak Ridge, Tennessee. The background signal from the empty sample holder is shown in grey. The arrows mark the position of the peaks in the density of states from low-energy acoustic phonons.
<u>Lower panel</u>: Peak energy in the phonon density-of-states, $E_{\text{ph}}$, vs doping $x$, obtained from neutron measurements on 12 different powder samples of Nd-LSCO. The energy of the peak in the acoustic phonon density of states increases by ~ 5% from $x = 0.12$ to $x = 0.40$, the full range of our specific heat study on Nd-LSCO. Given that the phonon specific heat goes as $\beta \sim E_{\text{ph}}^{-3}$, we expect a 7% decrease in $C_{\text{ph}}$ from $p = 0.16$ to $p = 0.24$, given the slight increase in $E_{\text{ph}}$.

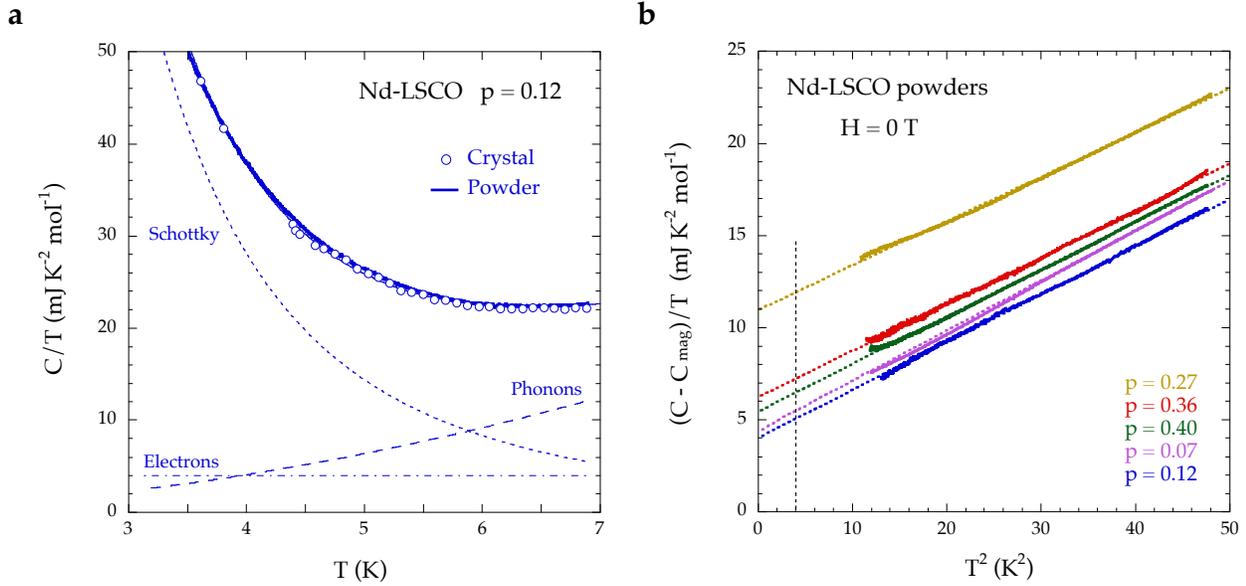

**Fig. S9 | Specific heat of our 5 polycrystalline samples of Nd-LSCO.**

**a)** $C / T$ vs $T$ for Nd-LSCO $p = 0.12$, comparing raw data on crystal and powder, as indicated. The solid line is a fit to the data, consisting of the sum of three contributions, plotted below: electrons (dash-dotted), phonons (dashed) and Schottky ($C_{mag} \sim T^{-2}$, dotted). **b)** Specific heat data for our powders with $p = 0.07$, 0.12, 0.27, 0.36 and 0.40, at $H = 0$, from which the Schottky term ($C_{mag}$) has been subtracted, plotted as $(C(H = 0) - C_{mag}) / T$ vs $T^2$. The dashed lines are linear fits to the data ($\gamma + \beta T^2$). For $p = 0.12$, the fit yields $\gamma \sim 4.0$ mJ / $K^2$ mol, in reasonable agreement with the value obtained by applying 18 T to suppress the Schottky anomaly in our Nd-LSCO crystal with $p = 0.12$ (Fig. 3b), namely $\gamma = 3.6 \pm 0.5$ mJ / $K^2$ mol (Fig. 4b). For the five powder samples, the $\gamma$ values are plotted in Fig. 4b (purple dots) and the $\beta$ values are plotted in Fig. S7d (orange dots). The value of $(C(H = 0) - C_{mag}) / T$ at $T = 2$ K is plotted in Fig. 4a (open red circles).

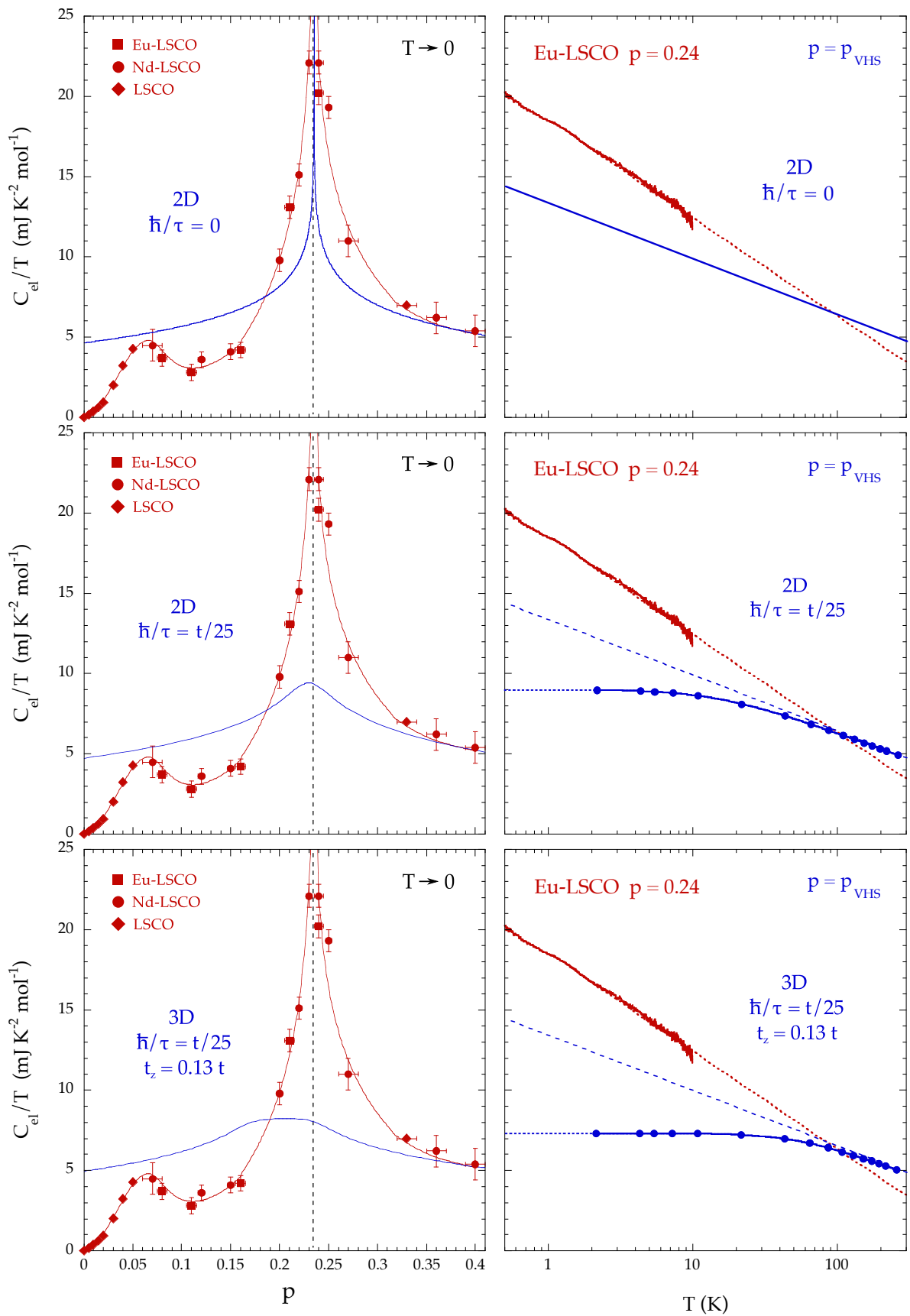

**Fig. S10 | Calculated specific heat from band structure and its van Hove singularity.**

Comparison of the measured specific heat of Nd-LSCO (red; 0.5 K data in Fig. 4b) and the specific heat calculated for the band structure of Nd-LSCO (blue; see Supplementary Text), with the van Hove point set to be at $p^*$. The calculations include the 3D dispersion in the Fermi surface (along $k_z$) and the disorder scattering, both consistent with the measured properties of our Nd-LSCO samples, namely their $a$-$c$ anisotropy in the conductivity and their residual resistivity (see Supplementary Text). We see that while the vHs can give rise to a cusp-like peak at $p_{vHs}$ and a log($1/T$) dependence of $C/T$ at $p_{vHs}$ in a perfectly 2D system with no disorder (top panels), these features inevitably disappear when the considerable 3D dispersion of the real material and the high disorder of the real samples are included (bottom panels). The calculations only quantify what is naturally expected: the rise in specific heat due to the vHs is cut off when $k_B T < \hbar \Gamma$, where $\Gamma$ is the scattering rate, or when $k_B T < t_z$, where $t_z$ is the $c$-axis hopping parameter. The fact that we see $C / T$ continuing to increase down to 0.5 K (lower right panel) completely excludes the vHs as the underlying mechanism.

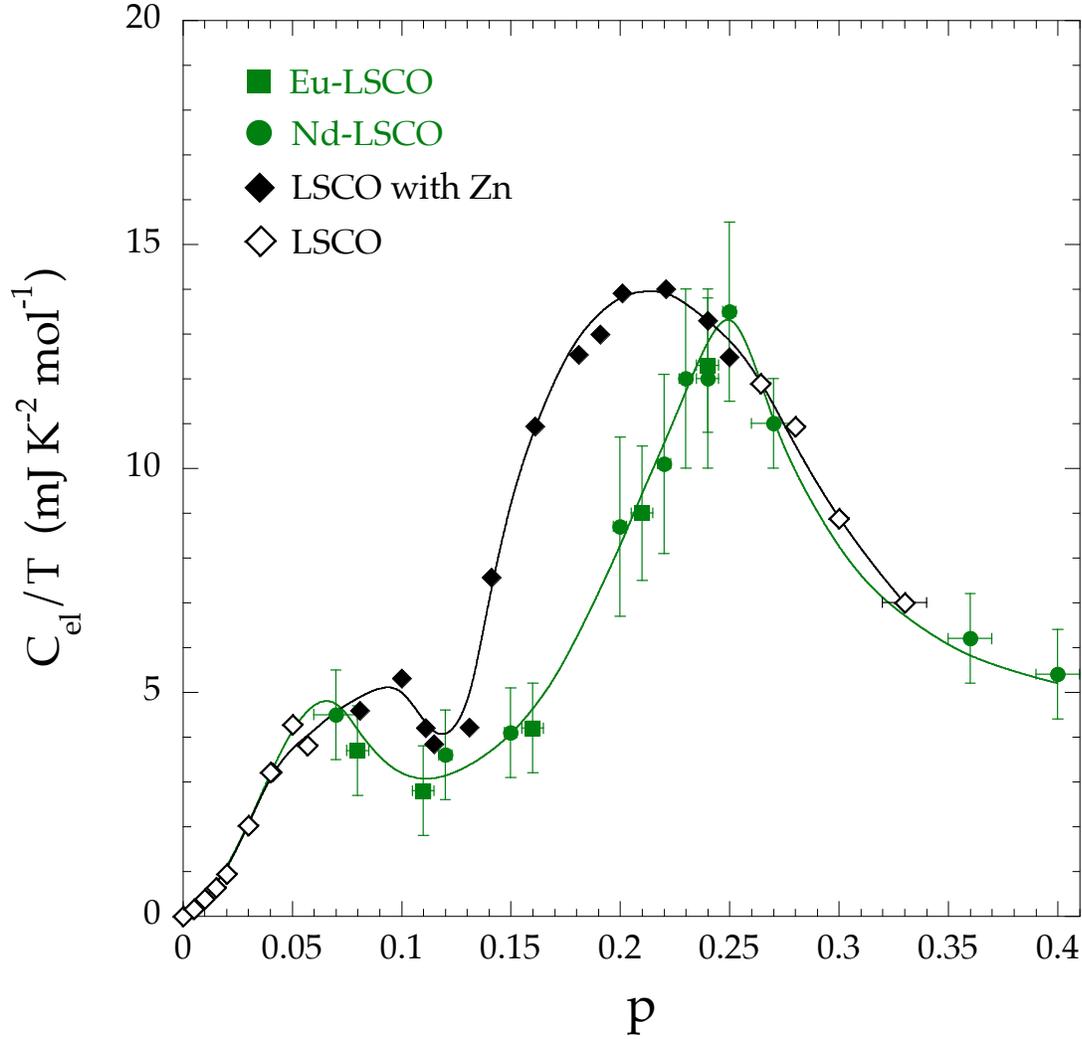

**Fig. S11 | Comparing with data on non-superconducting LSCO.**

Comparison of $C_{el} / T$ vs $p$ in our samples of Eu-LSCO and Nd-LSCO at $T = 10$ K (squares and circles, Fig. 4b) with published data on non-superconducting LSCO. Open diamonds are $\gamma$ measured in single crystals of LSCO at dopings where there is no superconductivity ($p = 0.33$ [16]; $p < 0.05$ [36]; remainder [21]). Full diamonds are data from powders made non-superconducting by Zn substitution [21]; $\gamma$ values are obtained from fits to $C / T = \gamma + \beta T^2$ between ~ 4 K and ~ 8 K. We see that these early data on LSCO are quantitatively consistent with our data on Eu-LSCO and Nd-LSCO, apart from a downward shift in the position of the peak, consistent with a lower $p^*$ in LSCO. All lines are a guide to the eye.

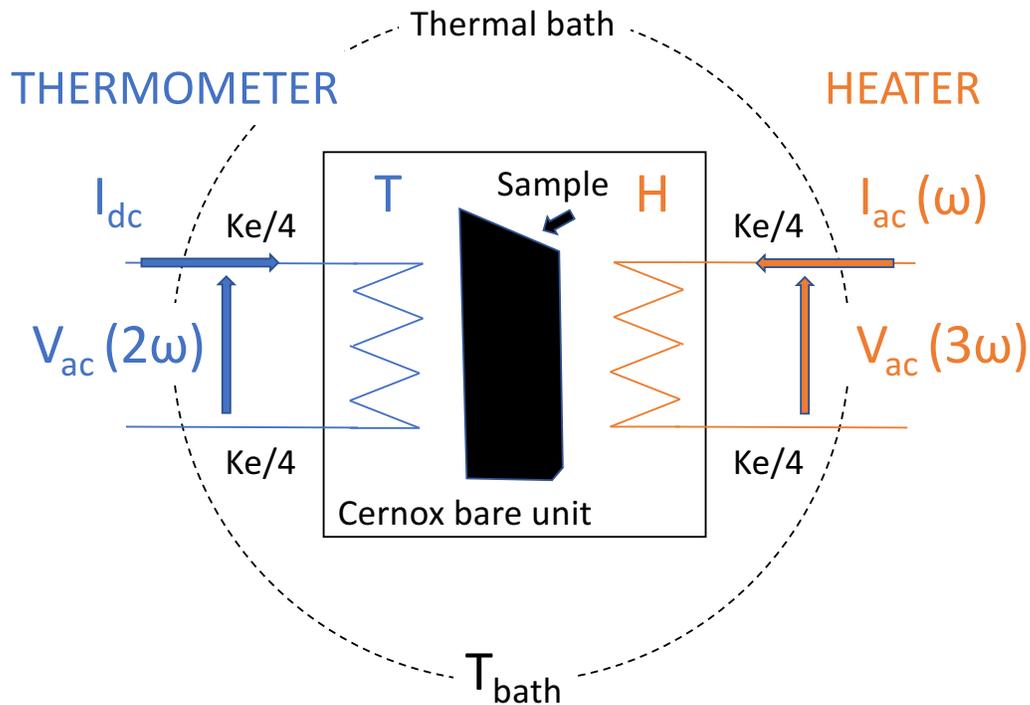

**Fig. S12 | Experimental setup for the measurement of heat capacity.**

Sketch of our experimental setup, showing the bare Cernox chip (black square) suspended by four PtW wires. A shallow groove is made with a wire saw to obtain two independent sides, one for the heater (H, right side) and one for the thermometer (T, left side). The sample is glued with a minute amount of Apiezon grease on the back of the sapphire substrate. An AC current $I_{ac}$ at a frequency $\omega$ is applied across the heater to induce temperature oscillations of the small platform (sample + Cernox). A DC current $I_{dc}$ is applied across the thermometer whose voltage is demodulated at $2\omega$ (see Materials and Methods – Specific heat measurement).

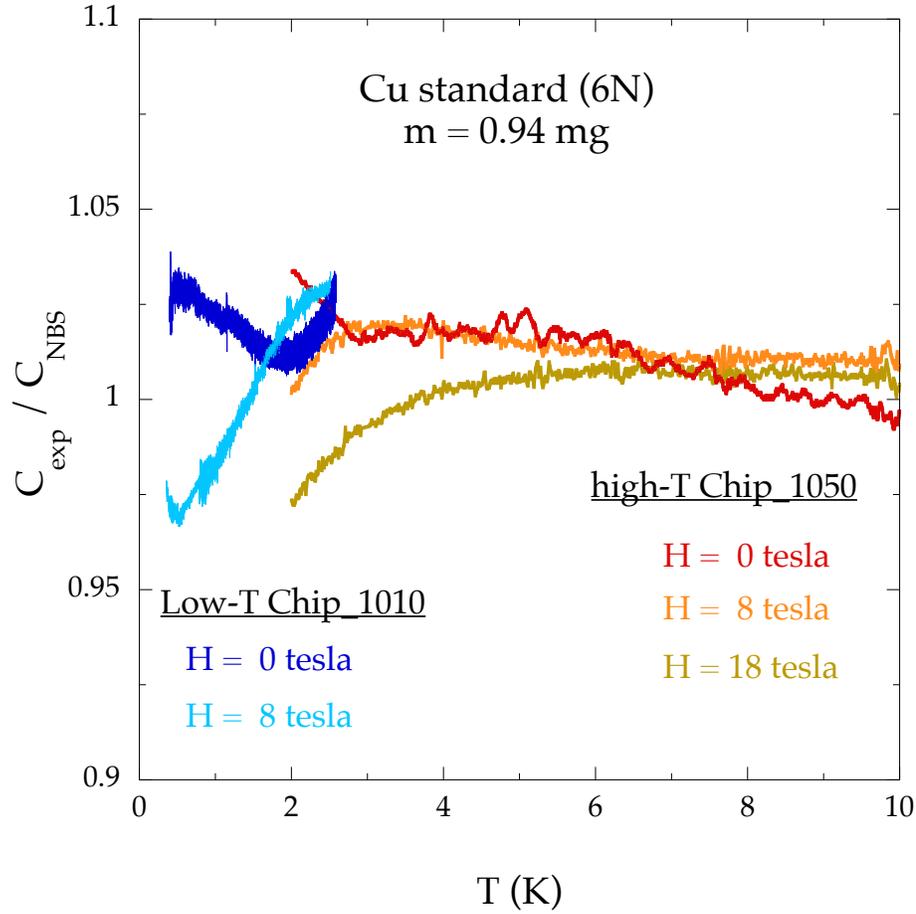

**Fig. S13 | Test of our specific heat measurement on a Cu sample.**

Specific heat $C_{exp}$ of a sample of copper measured using the same setup and analysis as used for our samples of Eu-LSCO and Nd-LSCO, plotted as $C_{exp} / C_{NBS}$ vs $T$, where $C_{NBS}$ is the standard value of the specific heat of copper established by the National Bureau of Standards. The measured data never deviate by more than 2-3 % from the standard, over the full temperature range from 0.5 K to 10 K, whether taken in the [4]He refrigerator at $H = 0$, 8 and 18 T (using a Cernox 1050 thermometer) or the [3]He refrigerator at $H = 0$ and 8 T (using a Cernox 1010 thermometer).